# Physical properties of thermoelectric cubic $La_{3-y}X_4$ lanthanum chalcogenides using first-principles calculations

RomainViennois, Kinga Niedziolka and Philippe Jund*

Institut Charles Gerhardt, Université Montpellier 2, Pl. E. Bataillon CC15003
34095 Montpellier, France

* corresponding author: pjund@univ-montp2.fr

Abstract:

We report ab-initio calculations of the stability, lattice dynamics, electronic and thermoelectric properties of cubic $La_{3-y}X_4$ ($X$=S,Se,Te) materials in view of analyzing their potential for thermoelectric applications. The lanthanum motions are strongly coupled to the tellurium motions in the telluride, whereas the motions of both types of atoms are decoupled in the sulfides. Nevertheless, this has no impact on their thermal properties because experimentally all compounds have low thermal conductivity. We believe that this is due to Umklapp scattering of the acoustical modes, notably by the low energy optical modes at about 7-8 meV found in all three chalcogenides, as in cage compounds such as skutterudites or clathrates, even though there are no cages in the cubic $Th_3P_4$ structure. We find that the energy bandgap increases from the telluride to the sulfide in good agreement with the experiments. However, due to their similar band structure, we find that all three compounds have almost identical thermoelectric properties. Our results agree qualitatively with the experiments, especially in the case of the telluride for which a great amount of data exists. All our results indicate that the sulfides have strong potential for thermoelectricity and could replace the tellurides if the charge carrier concentration is optimized. Finally, we predict also a larger maximum ZT for the p-type doped materials than for the n-type doped ones, even though compounds with p-doping have still to be synthesized. Thus our results indicate the possibility to make high temperature performing thermo-generators based only on $La_3X_4$ compounds.

## 1. Introduction

Chalcogenide rare-earths were studied for a long time because of their interesting physical properties such as superconductivity, mixed valences, strong electron correlations, magnetic, optical properties or thermoelectric properties. These properties depend on the stoichiometry of the compounds which have refractory properties and high stability until high temperature (especially for the sulfides) [1-8]. Among these compounds, we are interested by those crystallizing in the body-centered cubic crystal structure of $Th_3P_4$ type (space group I-43d) that are called the $\gamma$ phase inside the rare-earth-chalcogen phase diagram. They form a complete solid solution from $R_2X_3$ to $R_3X_4$ in the case of light rare-earth elements R [7-13]. In the case of the stoichiometric compound $La_3X_4$, a structure transition from the cubic form to a tetragonal form takes place and all three stoichiometric chalcogenide compounds are superconductors at low temperature [1,14].Even though the stoichiometric compounds $La_3X_4$ are bad metals (with about 4-6 $10^{21}$ $cm^{-3}$ electrons), the insertion of large amounts of vacancies gives rise to metal-semiconductor transition and the $La_{3-y}X_4$(with y about 1/3) or $La_2X_3$ compounds become very heavily doped semiconductors (with about $10^{20}$ $cm^{-3}$ electrons or less and energy bandgaps of about 1.5-3.2 eV depending of the chalcogen atoms) [1,6,7,10,14-46]. Consequently, their thermoelectric properties are very much improved and even more so since the presence of vacancies scatter strongly the acoustic phonons and therefore reduce strongly the thermal conductivity [20-42].

The thermoelectric properties of any materials are characterized by the Figure of Merit, $Z = \alpha^2\sigma / \kappa$, where $\alpha$ is the thermopower, $\sigma$ is the electrical conductivity and $\kappa$ is the thermal conductivity that is equal to the sum of the part coming from the lattice, $\kappa_l$, and of the part coming from the electrons, $\kappa_e$ [7,47]. This Figure of Merit has to be maximized in order to optimize any kind of thermoelectric material. For uncorrelated materials, the electronic properties are optimized in the case of heavily doped small bandgap semiconductors or semimetals with a small band overlap (with charge carrier concentration of about $10^{19}$-$10^{20}$ $cm^{-3}$). In this case the power factor (PF $=\alpha^2\sigma$) is maximized [7,47]. It is also necessary to minimize the thermal conductivity (mainly from the lattice) in order to keep a thermal gradient across the thermoelectric leg. To achieve that, it is necessary to scatter the full spectrum of the heat-carrying phonons (mainly the acoustic phonons) by introducing point defects or via alloying or doping in the case of short wave-length phonons. The same effect can be obtained by introducing inclusions of nanometric size or by reducing the grain size to few nanometers in the case of long wave-length phonons. Also, one can reduce the velocity of the heat-carrying phonons by the use of complex crystal structures or the presence of heavy atoms in the crystal structure [47]. Since the middle of the nineties and more particularly since the last five years, the search of new materials for high temperature applications is a very highly growing field notably because of the need to develop new sustainable and green energy sources due to the environmental problems and the stress on the energy resources. This is possible because of the development of new techniques of material synthesis, the discovery of new materials and the advances in theoretical concepts for the search of new materials for thermoelectric applications. In this framework, not only the search of new materials for thermoelectricity can be helpful but also the reexamination of old materials not sufficiently well studied such as the rare-earth chalcogenides, notably in light of other criteria such as the cost, abundance and toxicity of the materials and also their mechanical and thermal stability. These are indeed very important criteria in view of high temperature thermoelectric applications. There are actually two main problems in the development of thermoelectric materials: the small efficiency of the state-of-the art materials used in high temperature applications and the presence of rare and toxic elements such as tellurium (case of alloys based on $Bi_2Te_3$, PbTe, TAGS and LAST) or too expensive elements such as germanium (case of Si-Ge alloys and germanium clathrates) [7,47,48]. This is why we also need to develop new materials without tellurium or germanium.

In the past, the thermoelectric properties of the rare-earth chalcogenides have been the object of studies mainly from the sixties to the eighties and some conflicting results were reported showing the difficulties to optimize the thermoelectric properties [7,20-26,29,30]. By reviewing in a critical way the literature in 1988, Wood has shown that probably a ZT higher than 1 could be obtained for T > 1000-1200 K in $La_{3-y}Te_4$ and maybe also for rare-earth sulfides with compositions close to $R_2S_3$ (R = La, Pr or Dy) [7]. However, he pointed out that most of these results were reported with an extrapolation or even worse from an estimation of the thermal conductivity at high temperatures. This is due to the difficulty of the measurements at the highest temperatures, especially for the thermal conductivity and therefore the ZT values obtained for these compounds have to be taken with caution and can be subject to corrections [7].Therefore, Wood concluded that new and more accurate experiments on well characterized sample were needed to fully understand the potential of rare-earth chalcogenides [7]. Since that time, several groups have investigated rare-earth sulfides [31-37] and have confirmed that they have a relatively high ZT (about 0.7-0.8 at 1200 K for $RS_{1.48}$, the best composition according to refs. 30 and 35), but apparently lower than for lanthanum tellurides that have been recently thoroughly investigated by Snyder's group [39-42]. These authors found a large ZT of about 1.1 at 1275 K in the case of large amount of vacancies x and therefore confirmed the earlier studies done in the seventies on the TE properties of $La_{3-y}Te_4$ [7,22,24,26]. From the old studies [7,22,24,26] and the more recent studies of Snyder's group [39-42], it is now obvious that lanthanum tellurides have a large Figure of Merit, but a lot of work remains to be done to examine the potential for thermoelectric applications of the other compounds with $Th_3P_4$ structure. Actually, from the experimental data, it is difficult to understand if the tellurides have really significantly better TE properties than the sulfides or if this is due to a better optimization of the doping of the tellurides. Concerning the selenides, there are too few studies concerning the TE properties [7,26,37,38] and they don't permit to verify if they have TE properties comparable to the sulfides or the tellurides. Because of the problem of the low abundance of tellurium and even of selenium, it is obvious that if one could obtain a ZT higher than 1 in the rare-earth sulfides, this would have a large impact in the thermoelectric field.

Ab-initio calculations can be very helpful to understand the origin of the good TE properties of the tellurides, to compare them with those of the other rare-earth chalcogenides and to find the best way to optimize the TE properties of these materials. However, there are only few theoretical studies concerning the electronic properties of these rare-earth chalcogenides [11,40,42-46,49], and only some recent work of Snyder's group on the alloys based on telluride compounds were dealing with the thermoelectric properties of these compounds with doping [40,42]. There are no ab-initio calculations concerning their lattice dynamics or their stability. Therefore, the scope of the present paper is to report ab-initio calculations of the stability, lattice dynamics, electronic structure and thermoelectric properties of the three parent compounds $La_3X_4$ ($X$ = S, Se andTe) with the aim to analyze thepotential of these materials for thermoelectric applications, especially in the case of the sulfides that do not suffer of the problems of abundance and toxicity.

## 2. Computational details

First-principles calculations were performed using the projector augmented-wave (PAW) method [50,51] within the generalized gradient approximation (GGA), as implemented in the Vienna Ab initio Simulation Package (VASP) [52]. The calculations employed the Perdew-Bucke-Ernzerhof (PBE) exchange-correlation functional within the GGA [53]. We have used a plane-wave energy cutoff of 500eV held constant for all the calculations. For the relaxation of the structure in the primitive cell and the calculation of the equation of state, Brillouin zone integrations are performed using Monkhorst-Pack k-point meshes [54], with a k-point sampling of $15^3$and using the first order Methfessel-Paxton method [55] with a smearing of 0.2eV.The total energy is converged numerically to less than $1\times10^{-9}$eV/unit. After the structural optimization, the calculated forces are converged to less than $10^{-4}$eV/Å.For the electronic structure calculations, we have used the tetrahedron method with Blochl correction [56] and used the same energy criterion and k-point sampling than for the relaxation of structure. Charge transfers were calculated using the Bader Charge Analysis [57] with a k-points sampling of $30^3$. To ensure a high accuracy of the charge calculations we followed the recipe in [58] and tested the mesh for the augmentation charges starting from the mesh size used for the structural relaxations and increased it stepwise by 50% up to 350% . A grid size increase of 200% was enough to secure the convergence of the charge transfer between the atoms.

To determine the bulk modulus and its pressure derivative, we used the Vinet equation of state to fit the curve E=f(V) [59]. Lattice dynamics calculations were done using the frozen phonon method in the supercell approach as discussed by Parlinski [60]. With a k-point sampling of 3x3x3 we have calculated the Hellmann-Feynman forces in a relaxed 2x2x2 supercell of the conventional cellcontaining224 atoms with a precision better than $10^{-4}$eV/Åand subsequently the dynamical matrix has been diagonalized using Parlinski's Phonon code [60]. From these phonon calculations, the thermodynamic properties and the Atomic Displacement Parameter (ADP) tensors of each atomic type have been calculated (see ref. [60] for more details).

For the defect calculations, we have used the conventional cell and the primitive cell in which one lanthanum atom has been removed, leading to $La_{11}X_{16}$ and $La_5X_8$ respectively.

The transport properties (Seebeck coefficient) have been calculated using the BoltzTraP [61] program, with the Boltzmann Transport Equation (BTE) and the constant relaxation time approximation. The k-points sampling was fixed to $30^3$ as in the Bader charge calculations. Within the constant relaxation time approximation, the Seebeck coefficient $\alpha$ can be calculated directly and is not depending of the value of the relaxation time, contrary to the case of the electrical conductivity $\sigma$ and hence of the power factor PF=$\alpha^2\sigma$.

## 3. Results and discussion

### 3.1 Crystal structure and stability

The calculated values of the reduced position of the chalcogen atoms $x_X$, of the lattice constants a, of the formation enthalpies and of the bulk modulus and its pressure derivative are listed in Tables 1 and 2 together with the available experimental data [10-13,63-68]. Globally the calculated lattice constants are overestimated by at most 1.5% which is certainly due to the use of the GGA since it is well known that this approximation overestimates the lattice constants or the equilibrium volume [69].

The rare-earth chalcogenidesγ-$R_3X_4$crystallize in a body centered cubic cell with 2 formula-units per primitive unit-cell. In this structure, when the chalcogen coordinate has the ideal value of $x_X$= 1/12, there is only one type of La-$X$ bonding [10-13,62]. However, in our calculations as in the experiments with the best single-crystal refinements [10-13], a significant deviation of $x_X$ from 1/12 is found, meaning that there are two different kinds of La-$X$ bonds with different lengths. As can be seen from Table 1, it is interesting to note that both our calculations and the experiments [62] give for $La_3X_4$ an$x_X$ value of about 0.075 between 1/14 (=0.0714) and 1/12. Indeed, as discussed by Carter a long time ago, when $x_X$ = 1/14 it is possible to fill all the space in the $Th_3P_4$ structure by three Voronoi polyhedra: one corresponding to the rare-earth site and the two others being an enantiomorphic pair corresponding to two network of $X$ sites [62]. As can be seen in Carter's work based on electrostatic calculations, $x_X$ = 1/14 gives a more stable structure than the ideal $x_X$ = 1/12 [62]. Carter has performed these calculations in order to examine the possibility of vacancy ordering. It is thus interesting to note that in the case of $La_2X_3$, smaller $x_X$ values of about 0.0735-0.075 closer to $x_X$ = 1/14 were found experimentally [10,12,13]. Note that when making relaxation calculations based on the DFT for $La_3Te_4$, May et al found $x_X$ = 0.076 [42], a value slightly larger than in our calculations and the experiments.

The formation enthalpy of $La_3X_4$ ($X$ = S, Se or Te) in eV/atom can be calculated with the following equation:

$$\Delta H(La_3X_4) = E(La_3X_4) - (N_{La}E(La)/N_{tot} + N_X E(X)/N_{tot}) \qquad (1)$$

where $E(La_3X_4)$, $E(La)$ and $E(X)$ are the equilibrium first-principles calculated total energies (in eV/atom) of the corresponding $La_3X_4$ compound, of La with hcp ($P6_3/mmc$) structure, of S with face centered orthorhombic structure (F ddd) and of Se and Te with trigonal structure ($P\,3_121$), respectively. $N_{La}$ is the number of lanthanum atoms and $N_X$ the number of chalcogen atoms.

Concerning the formation enthalpy of $La_3X_4$, there are some recent experimental data for $La_3Se_4$, $La_2Se_3$ and $La_2S_3$ [64-66]. In addition Hepler and Singh [67] have discussed the available literature data and have mentioned that the data for $La_2Te_3$ have to be taken with caution. Therefore, we will compare our results with the results from refs. 64-66, except for the telluride for which one needs to be prudent. Our results concerning $La_3Se_4$ are in very good agreement with the experiments. When going from $La_3Se_4$ to $La_{11}Se_{16}$, we find that the formation energy only slightly decreases from -2.038 eV/atom to -2.056 eV/atom. This is in contrast with the experimental data for $La_3Se_4$ and $La_2Se_3$ for which also the formation energy is lower in absolute value (see tables I and II). This difference between our calculations and the experiment may be due to the other contributions to the formation enthalpy such as e. g. the vibrational contribution and/or also the fact that the $La_{11}S_{16}$ compound is an ordered vacancy phase whereas experimentally $La_2Se_3$ is a disordered phase. However, it cannot be excluded that the disagreement comes from the experimental side as the experimental data for the formation enthalpy values that are reported in Tables 1 and 2 come from different experiments and samples. Concerning the sulfides, our calculated formation energies for $La_3S_4$ to $La_{11}S_{16}$ are about 10 % smaller than the experimental value for $La_2S_3$.Our results are very close to the experimental results quoted in ref. 67 for the tellurides. Comparing our overall results with all the experimental data, we find quite high formation energies as in the experiments and the correct experimental tendency of a decreasing formation energy when going from the sulfide to the telluride.

We have also determined both the bulk modulus B and its pressure derivative dB/dP from the fit of the energy vs volume curve with the Vinet equation of state for all three stoichiometric compounds $La_3X_4$. The results are given in Table 1. As expected, the bulk modulus strongly decreases from the sulfide to the telluride. From a log-log plot of the bulk modulus vs average La-*X* bonding length $d_{La\text{-}X}$, we find that B decreases following approximately $d_{La\text{-}X}^{-3.67}$, which is very close to the expected dependence, which is B $\propto d^{-3.5}$ with d being the average bonding length [70]. We will discuss furthermore this result later in the section on the electronic structure. The agreement of our calculations with the experiment is excellent in the case of the $La_3S_4$ [68] and $La_3Te_4$ [39] compounds. There is no complete set of elastic constants data for $La_3Se_4$ but there are some data for other trivalent rare-earth selenides $R_3Se_4$ such as $Nd_3Se_4$ [71]. From the data obtained just above the magnetic transition (at about 50 K) for $Nd_3Se_4$, a bulk modulus B of about 55 GPa is found [71], a value smaller than from our calculations but larger than for the telluride, as in our calculations. Although due to the presence of non-vibrational contributions and of difference types of rare-earth atoms, one can conclude from the above comparison of our calculations with experimental data of $R_3X_4$ compounds that there is reasonable qualitative agreement.

The values we found for the bulk moduli of the rare-earth chalcogenides, i. e. 50 to 73.8 GPa, are quite similar than for other thermoelectric materials [72] for which the bulk modulus lies generally between 50 GPa (as for ZnSb) and about 90 GPa (as for the skutterudites).

## 3.2 Lattice dynamics and thermal properties

As $La_3X_4$ crystallizes in a body centred cubic structure with 2 $La_3X_4$ formula-units per primitive unit-cell, there are 42 different types of vibrational modes in the primitive unit-cell. At Γ point, these vibrational modes can be decomposed in irreducible modes as follows:

$$\Gamma_{vib} = \Gamma_{ac} + \Gamma_{opt} \quad (2)$$

with$\Gamma_{ac} = T_2$ and $\Gamma_{opt} = A_1 + 2\,A_2 + 3\,E + 5\,T_1 + 5\,T_2$.

Since the $A_1$, $E$ and $T_2$modes are Raman active, there are 9 Raman modes and since the $T_2$ modes are infrared active, there are 5 infrared modes. The $A_2$ and $T_1$ modes are optically silent.

We report the phonon dispersion curves, the total and partial phonon density of states of the three stoichiometric $La_3X_4$ compounds in the Figs. 1-3 (a,b). One can see that going from tellurides to sulfides, the motions of the lanthanum and chalcogen atoms become more and more decoupled. This is well illustrated by the partial density of states and can be highlighted by using the ratio between the cumulative spectral weight (CSW) of the lanthanum atoms and the chalcogen atoms $X$ (see Fig. 1-3 (c)), as already discussed in the case of other thermoelectrics such as the skutterudites [73]. Indeed, for the sulfide, one can see that this ratio reaches a very large value above 10 for an energy of about 8 meV before decreasing only for energies larger than 20 meV. This means that most of the lattice vibrations above 20 meV imply mainly the sulphur atoms whereas most of the lattice vibrations between 4 and 14 meV mainly imply lanthanum atoms. This behavior is less and less marked when going from sulfides to tellurides as illustrated by the reduction of the CSW ratio. In the case of the telluride, this ratio is always lower than 2, indicating that the motions of lanthanum and tellurium atoms are strongly coupled. The analysis of the behavior of the different vibrational modes as a function of the mass of the chalcogen atom also confirms this picture. Indeed, the energy of all the phonon modes with energy higher than 15 meV in $La_3S_4$ (excepted the $T_2$ modes) at Γ point scales with $1/(M_X)^{1/2}$.

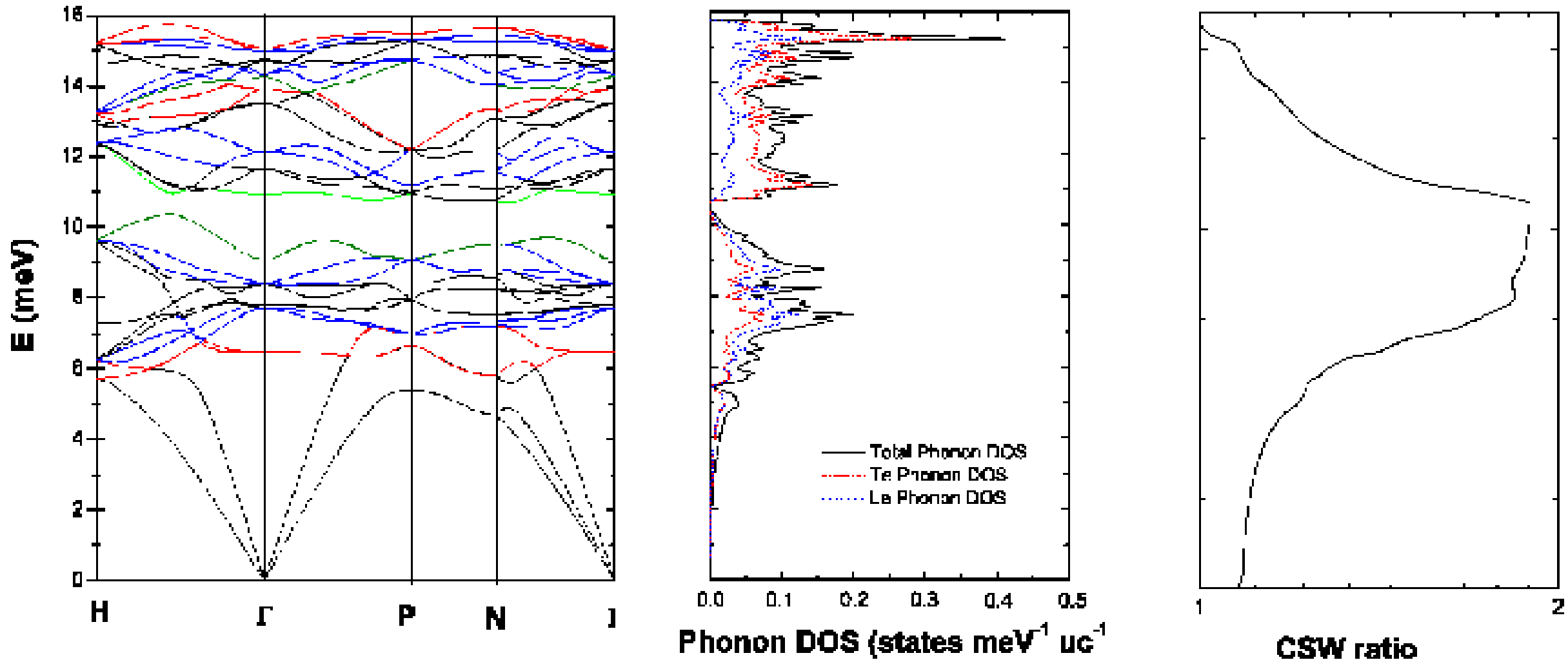

**Figure 1**

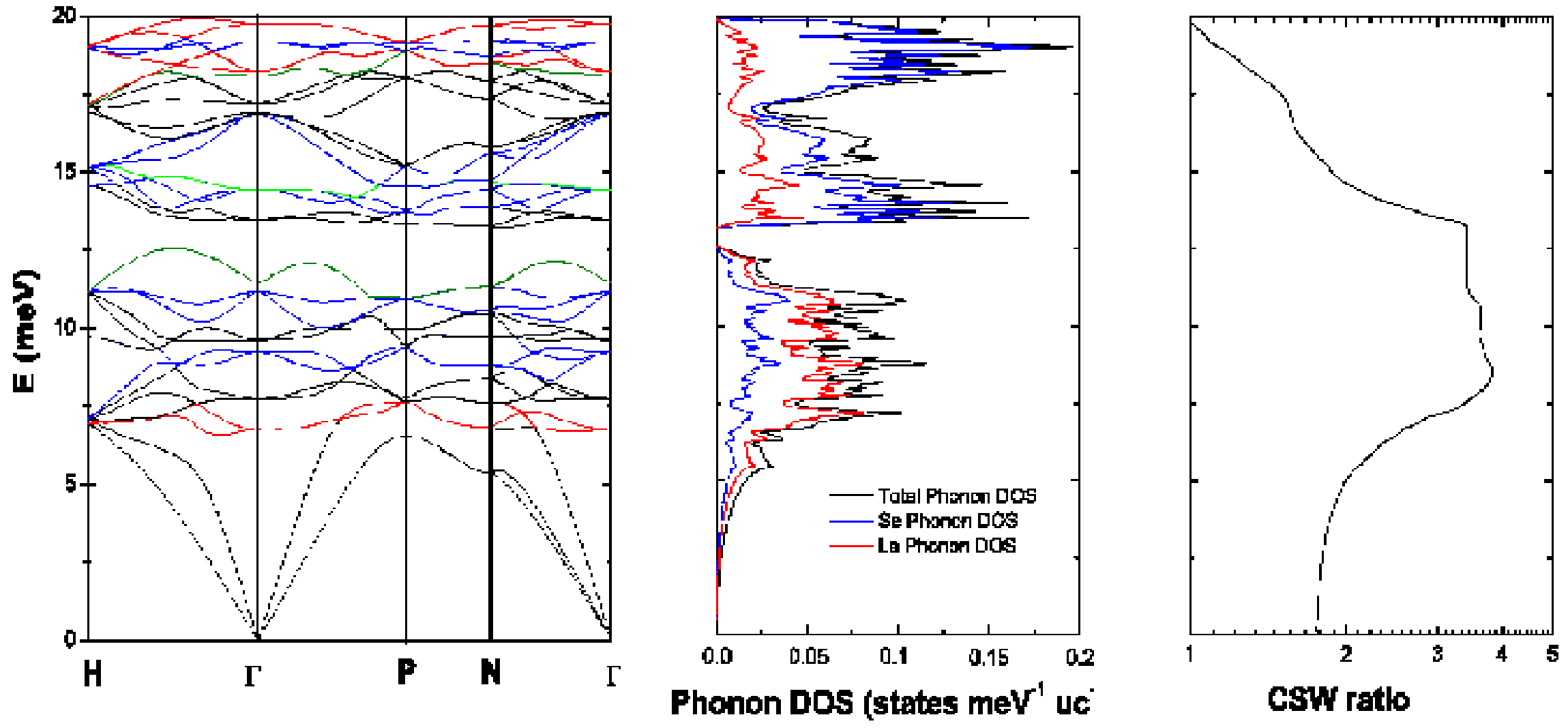

**Figure 2**

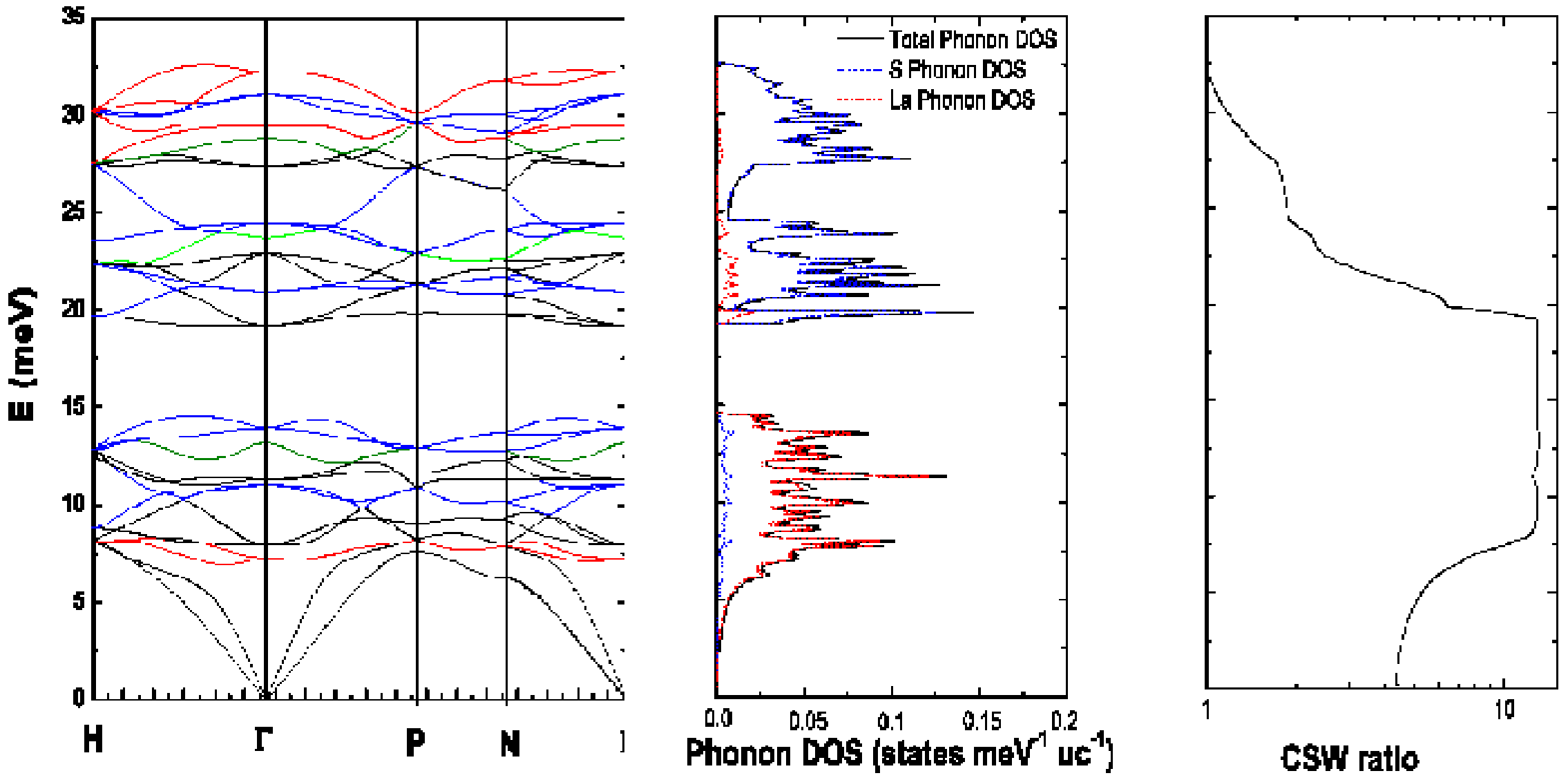

**Figure 3**

On the other hand, the energy of the lowest energy mode with $E$ symmetry scales with $1/(M_{av})^{1/2}$ (with $M_{av}$ = average atomic mass) and this is also the case for the lowest energy mode with $T_1$ symmetry (but the scaling is less satisfactory). Concerning the lowest energy modes with $T_1$ symmetry, the energy of the mode at about 7.5-8 meV remains constant in all three compounds, whereas the energy of the mode at 11.3 meV in $La_3S_4$ also scales with $1/(M_{av})^{1/2}$. Finally, we also note that one can see the presence of an energy bandgap in the phonon dispersion curves, below which the major contribution to the lattice vibrations comes from the heavier atoms (here the lanthanum) whereas above this energy bandgap, the major contribution tothe lattice vibrations comes from the lighter atoms (here the chalcogen atoms). This energy band gap becomes larger when going from the telluride to the sulfide and is certainly related to the decoupling between the motions of the heavy and light atoms mentioned earlier.

Now we compare our lattice dynamics calculations with the few available experimental data [41,74-79].In Table 3, we report the vibrational modes calculated at Γ point for all three compounds that can be compared with the experimental data from Raman and infrared spectroscopy available in the literature that we have reported in Table 4 [74-78]. In the case of $La_3Te_4$, the phonon density of states has been recently measured by Delaire et al [41] who found a broad maximum at about 8 meV, followed by a dip at about 10-11 meV, a shoulder at about 12 meV and another maximum at about 16 meV, the phonon DOS becoming zero at about 17-18 meV. These data agree well qualitatively with our results (see Fig. 3), as we find a maximum at 7-8 meV, followed by a gap at about 10-11 meV that corresponds well to the dip experimentally observed, then we observe a small maximum slightly above 11 meV and a large maximum at about 15-16 meV. The phonon DOS becomes zero at about 15-16 meV, below the experimental value. This overall qualitative agreement with the experiment makes us confident about the quality of our calculations for the two other compounds. We note that Delaire et al have observed that the vacancies have an effect not only on the broadening of the phonon DOS but also to up-shift the structure observed in the phonon DOS [41]. Indeed, the low energy peak shifts from 8 to 9 meV and the phonon DOS becomes zero at 20 meV, a significantly larger energy than for the stoichiometric compound. Note also that the dip at about 10 meV is partially filled in this case. We point out that previously the same kind of effect has been observed by Raman scattering experiments for the sulfides [75]. Indeed, an increase of the higher energy Raman mode from 260 $cm^{-1}$ to 280 $cm^{-1}$ was also observed in the case of $La_{3-y}S_4$ compounds when y is increasing from y = 0.33 (corresponding to cubic $\gamma$-$La_2S_3$) to 0.22 [75]. Note that the same observation can be done for alkaline-earth substituted $La_2AS_4$(A = alkaline-earth) [74]. This phenomenon seems therefore a general behavior in these compounds and calls for further investigation in a more systematic way.

Unfortunately, the determination of the mode symmetry has been performed with polarized Raman experiment only in the case of the cubic $\gamma$-$La_2S_3$, i. e. for the compound with the vacancies [77]. This will make more difficult a quantitative comparison with our calculations due to the changes induced by the defects in the vibrational spectrum, as discussed above. Nonetheless, we try now to compare these experimental data in resonant conditions (reported in Table 4) with our calculations done for the fully stoichiometric compound, $La_3S_4$. Also, in this table are reported the data of unpolarized Raman experiments [76] and infrared experiments [78] for the cubic $\gamma$-$La_2S_3$. Our calculations agree very well with the experiments for the $A_1$ mode. Concerning the $E$ modes, the agreement is very good for the mode at about 22 meV, but it underestimates of about 15 % the high energy $E$ mode and even more for the low energy $E$ mode. Also for the $T_2$ modes, the low and high energy modes are strongly underestimated. This underestimation of the energy mode at low and high energy could be related to the presence of vacancies as discussed previously. Indeed, for lower vacancy concentrations, the 280 $cm^{-1}$ (about 35 meV) Raman mode decreases to 260 $cm^{-1}$ (about 32 meV) [75], which is much closer to our calculation results. However, we note a good agreement of our results concerning the three TO modes of $T_2$ symmetry observed in the infrared experiment on $LaS_{1.49}$ reported by Ivanchenko and co-workers [78]. Obviously, as our calculations are dealing with the metallic stoichiometric compounds, one cannot conclude anything about the LO modes.

We note that experimentally there is an uncertainty for the position of the other $T_2$ modes, excepted for the mode at about 22-23 meV that was found in all experimental reports [76-78]. In this region of the spectrum, we note that Koselov et al also suggest the presence of a second $T_2$ mode at about 20 meV [77]. Our calculations seem to confirm this assignment as we have found the presence of two $T_2$ modes at 19.2 and 22.9 meV. However, we do not confirm the other proposal of Koselov et al [77] about the last $T_2$ mode that they propose to be at about 33.5 meV. Instead we find that the last $T_2$ mode is around 11.3 meV. It could maybe have a low Raman activity explaining why it is difficult to observe it. Another possibility is that it could correspond to the small broad peak at about 15.5 meV found by Knight and White [75]. We also note the presence of a very weak and broad peak at about this energy in the polarized non-resonant spectra for $A_1+E$ and $T_2$ symmetry modes, in unpolarized non-resonant Raman spectrum and it is even more evident in all polarized resonant spectra in the paper of Koselov et al [74]. However, we note that in these last conditions, as also discussed by Koselov et al [74], the Raman spectroscopy is more sensitive to the presence of defects and we can therefore not exclude the Raman activation of silent modes by the presence of the vacancies that could relax the Raman selection rules. Our calculations indicate the presence of two silent modes at about 13.5 meV that could maybe explain these possible defect induced-Raman peaks.Finally, we want to note that both our calculations and the polarized Raman experiments on $La_2S_3$ do not confirm the presence of a TO infrared mode at about 26.5 meV as determined by Ivanchenko et al [78], neither the presence of a low energy mode of $T_2$ symmetry at about 4.7 meV in $CaLa_2S_4$ found by Merzbacher et al [79]. Indeed, we do not find any optical mode below 7 meV for the sulfide, whereas the position of the 4 other modes observed in their infrared experiments match well with our calculations, although slightly shifted to 7.4, 10.7 17.8 and 25.5 meV, which is not surprising as one-third of the lanthanum are substituted by calcium. Therefore, the low-energy mode found in IR experiments on $CaLa_2S_4$ must originate from defects induced by the presence of calcium.As seen above, actually, only few experimental data are available on the lattice dynamics of these chalcogenides, essentially for off-stoichiometric or alloyed compounds, and our results call for new experiments in this field.

In the case of thermoelectric materials, the Atomic Displacement Parameters (ADPs) $U_{ij}$ have proved to be efficient parameters for studying the dynamics of atoms and have been often connected to low-energy modes [73,80-82]. We have also calculated such parameters in the case of the $La_3X_4$ compounds. The isotropic and anisotropic ADPs are reported in Table 5. The largest ADPs were found for the tellurides for both the lanthanum and the tellurium atoms at all temperatures and approach 0.01-0.013 $Å^2$ at room temperature. These results agree reasonably well with the experiments (see Tables 6) [10,12,13], although the experimental data in the literature are only for the compounds containing vacancies, i. e. the $La_2X_3$ compounds. Note that the best agreement for the isotropic ADPs is with the most recent experiments performed on the $La_2Se_3$. These values are relatively large, as in ZnSb [72], but smaller than in the case of intercalated atoms in skutterudites or clathrates where they approach about 0.02 to 0.03 $Å^2$ at room temperature [73,80-82]. These large ADPs are related to the low energy modes present in these compounds. Indeed, it is possible to fit the temperature dependence of the ADP of the lanthanum atoms using a simple Einstein model with a very small disagreement below 40-50 K. This works very well and we find 102 K (8.8 meV), 97.2 K (8.38 meV) and 91.2 K (7.86 meV) for respectively the Einstein temperature of lanthanum in $La_3S_4$, $La_3Se_4$ and $La_3Te_4$. It is also interesting to note that the ADP of the chalcogen atoms is the largest for the sulphur atoms below about 150 K and becomes the smallest above about 250 K. Conversely, the ADP of the tellurium is the smallest below about 75 K and becomes the largest above about 250 K. This can be explained by the larger zero point motion of the sulphur at 0 K and its larger Debye temperature above 0 K. Indeed, the high temperature slope is inversely proportional to the Debye or Einstein temperature of the atoms. When fitting the ADP of the chalcogen atoms using an Einstein model, we find 102.7 K (8.85 meV), 135.2 K (11.66 meV) and 217.5 K (18.75 meV) for respectively the tellurium, the selenium and the sulphur. Interestingly, these Einstein temperatures scale with $1/(M_X)^{1/2}$ as expected. The above observations confirm that the motions of the tellurium and lanthanum atoms in $La_3Te_4$ are strongly coupled, whereas the motions of lanthanum and sulfur atoms are strongly decoupled in $La_3S_4$, $La_3Se_4$ having an intermediate behavior.

Now, we describe our results concerning the thermodynamic properties of the $La_3X_4$ compounds. Our results for the heat capacity are reported in the inset of Fig. 4.

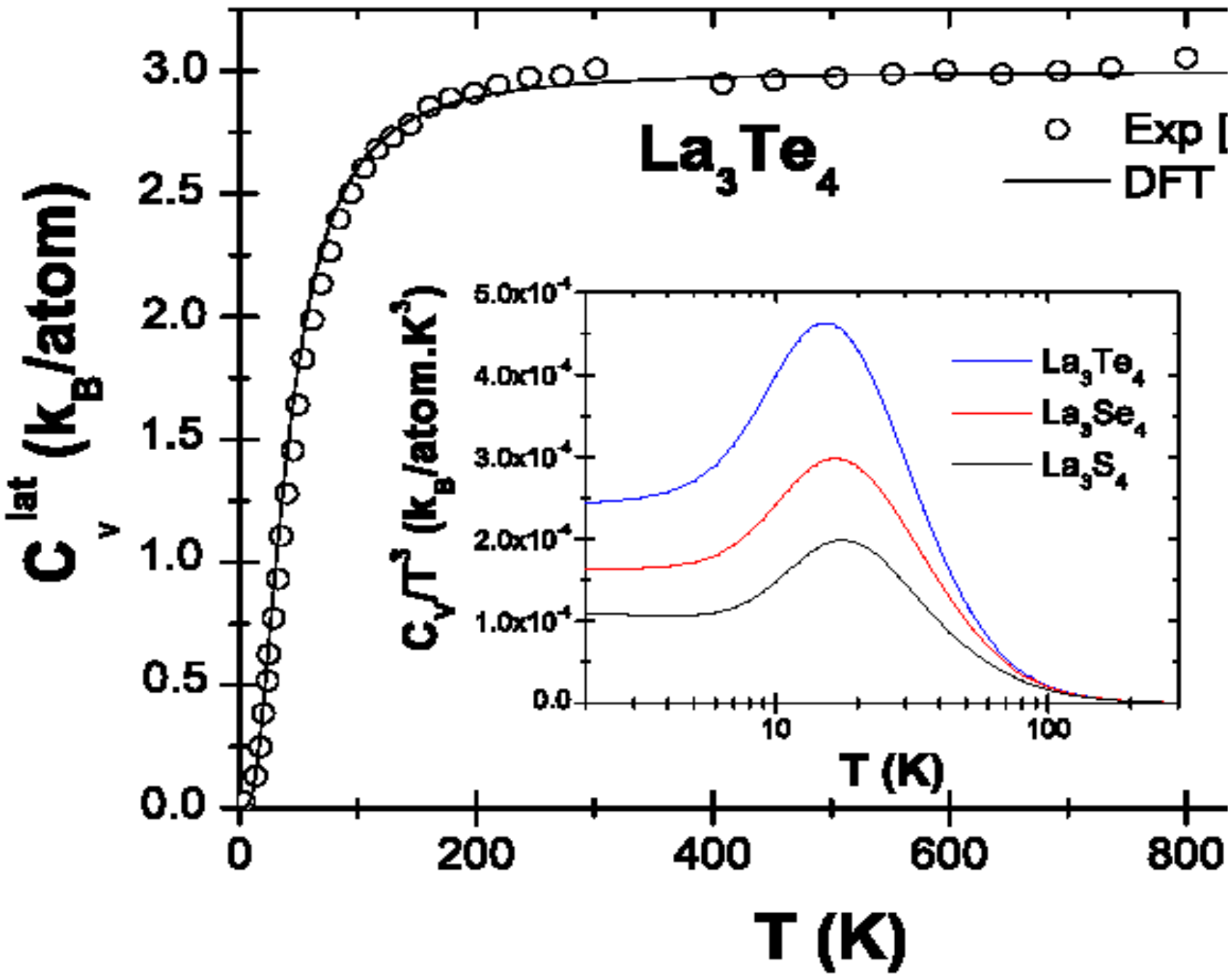

**Figure 4**

Our calculations are able to reproduce well the heat capacity measured by Delaire et al [41], as can be seen in Fig. 4 where our calculations are compared with Delaire's data with the electronic part subtracted from their data. In Table 7, we also show the comparison of our calculated heat capacity with data from the literature for $La_3Se_4$ [64] and $La_3S_4$ [8]. The agreement is less good than with the Delaire's data, but this is certainly due to the lower accuracy of the experimental data in refs. 8 and 64 because in these papers the heat capacity at 298 K is determined from a fit of relatively scattered experimental enthalpy data.

In the inset of Fig. 4 where we show a $C_V/T^3$vs T plot in a semi-logarithmic scale for the different compounds, one can see a maximum whose temperature increases from 15 to 18 K when going from the telluride to the sulfide. These maxima correspond to the features in the phonon density of states below 10 meV. Indeed, when plotting $G(E)/E^2$vsE (with G(E) being the phonon DOS), one finds a maximum at 7.5 meV in the telluride whose energy increases to about 8 meV for the sulfide. However, it is not possible to attribute the broad maximum in the plot $C_V/T^3$vs T plot to a specific mode as its origin comes from all the modes in this energy range. It is interesting to note that these values are very close to those obtained with a simple Einstein model used previously to fit the ADP of the Lanthanum atoms.

From our lattice dynamic calculations, we have calculated the Debye temperature $\theta_D$ by using different methods such as the low temperature heat capacity $\Theta_D{}^C$ and the integration of the phonon DOS $\Theta_D{}^{int}$in order to compare our results with available experimental results determined by Delaire et al [41] in the same manner. The agreement is quite good for the telluride but less good for the selenide and sulfide for which the Debye temperature is overestimated by about 10-15 %.

From our ab initio calculations we can evaluate the thermodynamic Grüneisen parameter $\Gamma$and thus the volume thermal expansion $\alpha_V$ (because we have already determined $B_M$ and $C_V$) by using a relation implying dB/dP determined by fitting the energy vs volume curve with the Vinet equation of state (see above). Now, we can use the Dugdale and McDonald approximation as follows [83]:

$$\Gamma^{DM} = -1/2 + (1/2)dB/dP \quad (3)$$

This way, we find $\Gamma^{DM}$ = 2.1 for $La_3S_4$ and $\Gamma^{DM}$ = 1.9 for $La_3Se_4$ and $La_3Te_4$ as dB/dP = 5.2 and 4.8 respectively for these cases. These values are larger than the experimental determination for $La_3S_4$and $La_3Te_4$ for which it was found respectively $\Gamma$= 1.32 [68] and 1.76 [41] for the thermodynamic Grüneisen parameter. The agreement is satisfactory, especially for the telluride, given the approximations used. Note however that Fütterer et al [68] were able to determine the Grüneisen parameter from the acoustical modes only, $\Gamma_{elast}$, for the case of $La_3S_4$ and they found a value significantly higher of about 2.85.

In the next step, we aim to estimate the thermal conductivity, κ, using a very simple model considering only the Umklapp scattering in order to see if this mechanism can be the dominant mechanism of the phonon scattering. This is justified because the lattice thermal conductivity of the $La_{3-y}X_4$ compounds with the lowest charge carrier concentration has been observed to decrease with increasing temperature above room temperature [35,39,84], as expected for Umklapp scattering (see below). Several authors have discussed the validity of different formulations of the Umklapp scattering contribution to the thermal conductivity in a general manner. Following Slack [85], for complex structures, it is necessary to use:

$$\kappa_l = A\, M_{at}\, (V_{at})^{1/3} \Theta_D{}^3 / T(n^{1/3}\Gamma)^2 \qquad (4)$$

where $M_{at}$ is the average atomic mass, $V_{at}$ is the volume per atom, $\Theta_D$ is the Debye temperature, A is a constant equal to $3.04*10^{-8}$ $s^{-3}K^{-3}$, n is the number of atoms in the primitive cell and $\Gamma$ is the Grüneisen parameter. In the above formula, the thermal conductivity is obtained in W/cm.K if the volume is given in $A^3$ and the average atomic mass in amu. If we want to determine the thermal conductivity from our calculations, we need to determine the Grüneisen parameter. Thus, we use the Grüneisen parameter estimated from Eq. 3 and the dB/dP value found from the fit of the Equation of State with the Vinet formula, as we have previously already done in the case of ZnSb [72].

Using these values of $\Gamma$ and the Debye temperature calculated from the heat capacity at low temperature$\Theta_D^C$, we find $\kappa_l$ = 1.63, 1.8 and 1.61 W/m.K for respectively $La_3S_4$, $La_3Se_4$ and $La_3Te_4$ at 300 K. The experimental values of the lattice thermal conductivity for the stoichiometric compounds are respectively 0.8-2 [23,24,32,34,36], 1 (for $Gd_3Se_4$ and $Nd_3Se_4$) [38] and 0.5-1.7 [23,24,39] W/m.K.The agreement with the experiments is fair, taking into account all the approximations used in our calculations as well as the scattering of the experiment results. Using the elastic Grüneisen parameter $\Gamma_{elast}$ found by Fütterer [68] together with the Debye temperature determined for $La_3S_4$ from the heat capacity (227 K) [14] and the experimental value of the volume $V_{at}$ (see Table 1), we can calculate the thermal conductivity using (4) and we find 0.57 W/m.K, a value about 3 times lower than when using only the calculated values and about 2 times lower than the experimental values.

From the above discussion, one can see that the Umklapp processes are able to explain the origin of the low thermal conductivity in the lanthanum chalcogenides. This result has a general interest for the search of new thermoelectric materials with low thermal conductivity because these compounds have low energy optical modes but do not contain cages in their structure. This result supports the Umklapp scenario for lots of cage compounds among the skutterudites and clathrates [73,86,87] because it shows its high efficiency to reduce the lattice thermal conductivity in compounds containing heavy atoms but not intercalated in any cage and having low-energy optical modes. In this aspect, these compounds are closer to lead telluride, but with a much smaller anharmonicity than this last one although PbTe has a larger lattice thermal conductivity (about 2 W/m.K) [88]compared to the $R_3X_4$ compounds.

It is worth mentioning that some alternative scenarios implying a strong hybridization between acoustical and optical modes were proposed to explain the low thermal conductivity of these cage compounds [89,90].

### 3.3 Electronic properties and bonding

As can be seen in Figs. 5-7, the stoichiometric compounds are metals with the Fermi level close to a peak in the density of states.

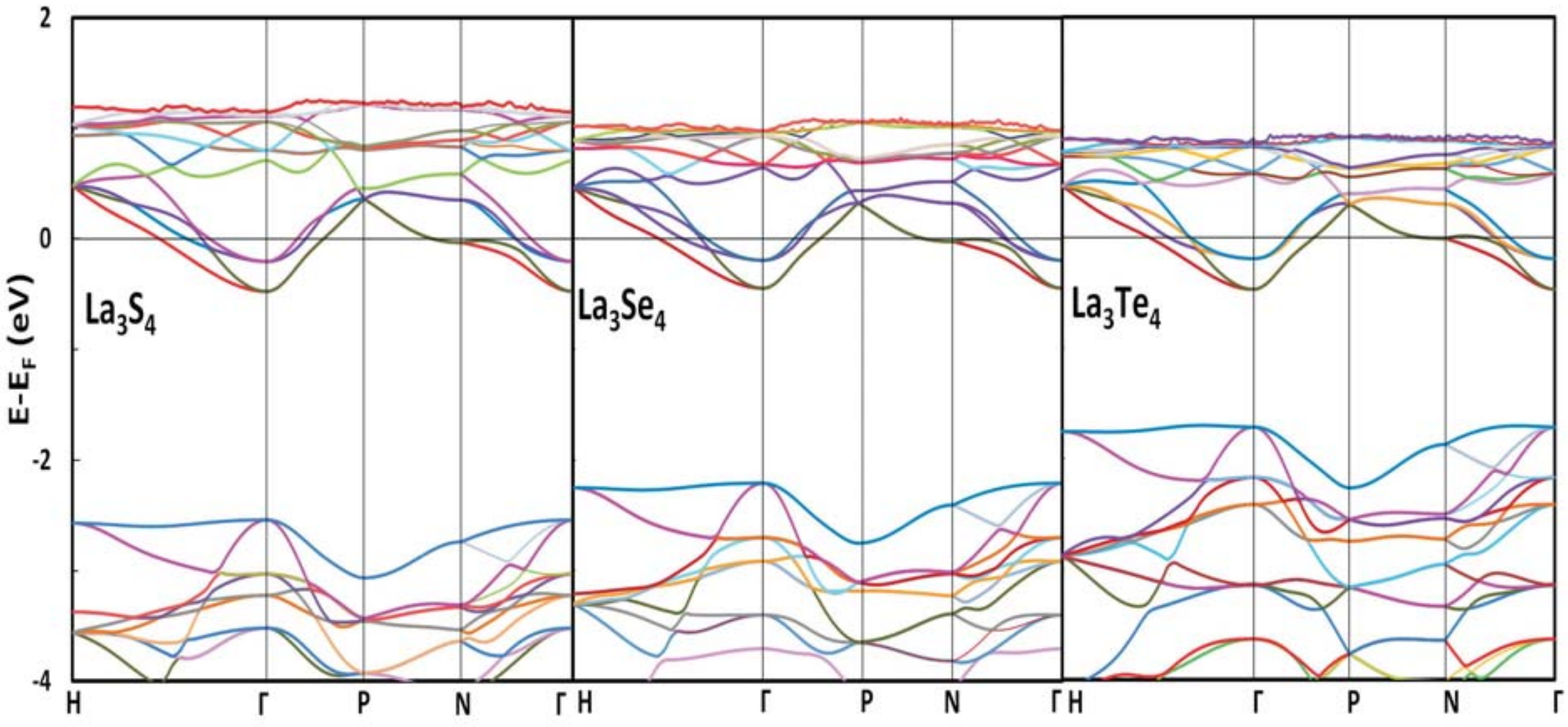

**Figure 5**

The proximity of the Fermi level with this peak explains partly why these materials become superconducting at low T. We have calculated the Sommerfeld coefficient, γ, from the density of states at the Fermi level $N(E_F)$ and found good agreement with the Sommerfeld coefficient in the experiments [14,41] (see Table 8). Note that we compare our data with the values estimated by Westerholt for the cubic phases of stoichiometric compounds [14]. If we want to compare our data directly with the experimental data of the cubic superconducting phases, we have to extrapolate our results to the $La_{2.974}S_4$ and $La_{2.985}Se_4$ compounds in shifting the Fermi level by assuming a rigid band approximation and by assuming that each Lanthanum vacancy removes 3 electrons. We will discuss later how valid this rigid band approximation is. As observed experimentally, we also find larger γ for the sulfide and the selenide than for the telluride and this explains why the highest superconducting transition temperature $T_{sc}$ was found for the $La_3S_4$ and $La_3Se_4$ compounds.

From Figs. 5-7, one can see that there is an energy bandgap located at about 0.5 eV below the Fermi level and it increases from the telluride to the sulfide. For the selenide and sulfide, the energy bandgap is direct but it is worth mentioning that the top of the valence band is only slightly higher in energy at the Γ point than at a point in the Γ-H direction (in fact it is only about 20 meV higher). In the case of the telluride we find even that the energy bandgap is indirect. In all cases, this observation together with the flatness of the highest valence band points an interesting potential for thermoelectric applications in n-doped alloys derived from $La_3X_4$ compounds as it is well known that band degeneracy and large effective masses lead to an increase of the thermopower (see e. g. refs. 7 and 47). We will come back to this point later when we will discuss our results on thermoelectric properties.

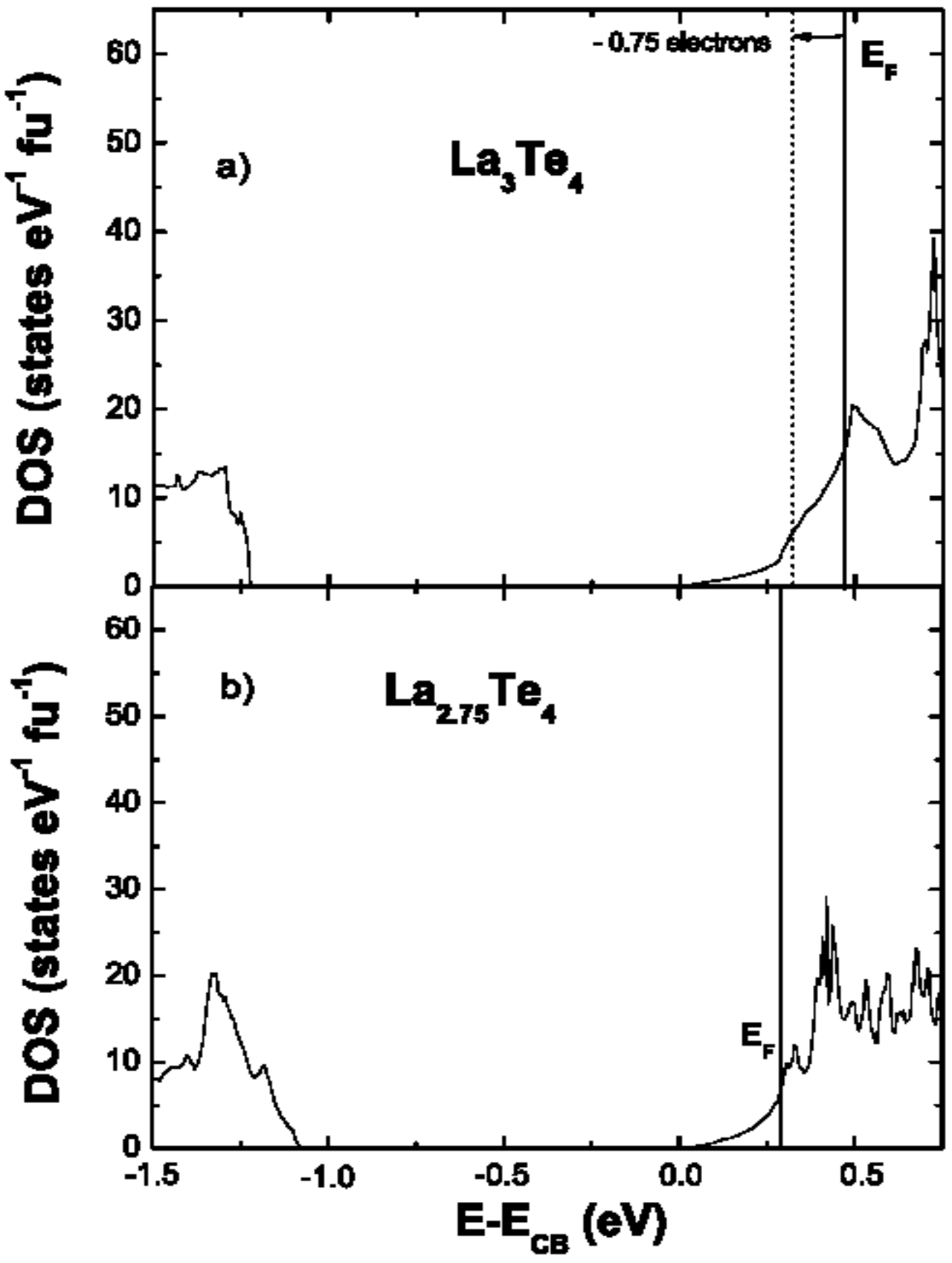

**Figure 6**

Note that as usually in DFT calculations, we find smaller bandgaps than in experiments. As only the compound with $La_2X_3$ composition is semiconducting, we have to compare our results for $La_3X_4$ composition with the experimental values from optical experiments on these $La_2X_3$ compounds. We find bandgap energies of 1.23 eV, 1.76 eV and 2.06 eV for respectively $X$= Te, Se and S, to compare to bandgap energies of about 2.2 eV [17], 2.6 eV [18] and 2.4-2.9 eV (with 2.8-2.9 eV being the most likely value) [15,16,19,91] for respectively $X$ = Te, Se and S in the optical experiments. In all these experiments, the authors have modeled their absorption data by assuming that the energy bandgap was direct, except in ref. 15.As also discussed in this last reference, we note that the experimental situation is not very clear and it is difficult to assess that the energy bandgap is clearly of direct nature without doubt. This could be explained either by the complex band structure of these compounds or also by the presence of defects because these experiments were carried out on the semiconductor $La_2X_3$ compounds that contain a great amount of vacancies. Indeed, looking at the band structure, it is possible to see both direct and indirect optical transitions, making difficult the interpretation of the absorption experiments solely from their wave-length dependence. Note also that some other processes as e. g. the excitonic transitions could also complicate the interpretation of the optical spectra.

Only few DFT calculations were reported in the past and never for all three $La_3X_4$ compounds for comparison. Zhukov et al have used tight-binding LMTO-ASA technique in order to study both $R_3S_4$ and $R_2S_3$ compositions (with R = La, Ce) and they found an energy bandgap of about 2.3 eV in the case of the lanthanum compounds [43]. They found an indirect energy bandgap between Γ point and H point. We note also that they found that the Fermi level is shifting in the valence band for the $R_2S_3$ composition, in agreement with our finding for $La_5S_8$composition (see below). The LMTO-ASA technique was also used by C. Felserfor the sulfide but within the DFT and using the LDA exchange-correlation functional [44]. She found a wide bandgap of about 2.5 eV located at about 0.5 eV below the Fermi level. The $R_3S_4$ compounds (with R = La, Ce) were studied by Shim et al [45] and Kang et al [46] using LSDA and LMTO type calculations and they found a wide band gap of about 3 eV, a value close to the experiment. However, no details were given for the calculations. The features found in the XPS spectra determined by Kang et al [46] for $La_3S_4$ than and $La_3Se_4$ agree reasonably well with our calculations, notably the presence of a gap of about 2 eV that appears at about 0.5-1 eV below the Fermi level. The authors found a valence band width of about 4-5 eV, to compare with about 4 eV in our calculations. Note however that these XPS spectra are not accurate enough to be able to make a more detailed comparison with calculated band structures.

More recently, May et al have performed electronic structure calculations using a pseudopotential DFT code with the PBE exchange-correlation functional [42] and found similar energy bandgaps for both $La_3Te_4$ and $La_3S_4$ than us. They found however a larger bandgap than in their previous calculations for $La_3Te_4$ where they have used $x_X = 1/12$ instead of relaxing this structural parameter [40]. As in our calculations, they found that the highest valence band is very flat, making it difficult to know if the energy bandgap is direct or not. Thus further work both experimentally and theoretically is necessary before to conclude if the energy bandgap is direct or not in these compounds.

We have calculated the Bader charges of the atoms for the stoichiometric compound $La_3X_4$ (see Table 9) and found that it is smaller than in the pure ionic case, which is not surprising giving that these compounds are bad metals with about 1 electron/f. u.. However, the charges are still relatively well localized on the atoms and this means that the bonds have still some significant ionic character, which is increasing from the telluride to the sulfide compounds.

In order to test the band rigid approximation (RBA), we have performed calculations of the electronic structure of $La_{11}X_{16}$ and $La_5X_8$ in their ordered form and find that the Fermi level in these compounds is the one of $La_3X_4$ shifted by an energy corresponding to the reduction of the number of electrons: respectively 1.5 and 3 electrons per primitive body-centered unit-cell (see the exemple of $La_{11}X_{16}$ in Fig. 6).We find that the Fermi level is still in the conduction band in the case of $La_{11}X_{16}$, whereas it is inside the conduction band in the case of $La_5X_8$. Our results are in good agreement with some previous calculations in the literature [40,42,43] and confirm the validity of the use of the RBA for $La_{3-y}X_4$ compounds in order to study their electronic and thermoelectric properties. Therefore, one can see the lanthanum vacancies as electron acceptors, as it has been observed experimentally.

### 3.4 Thermoelectric properties

We have calculated the thermoelectric properties of the three stoichiometric compounds $La_3X_4$ by means of the Boltzmann Transport Equation approach [61,92-94]. In Fig. 7, we report how the power factor (PF) is changing with the position of the chemical potential at 300 K and 1273 K. This PF has been determined with the assumption that the relaxation time $\tau$ is equal to $2*10^{-15}$ s. This value has been chosen since it gives the best agreement between the available experimental data of the electrical conductivity for $La_3Te_4$ and $La_3S_4$ and the calculated electrical conductivities as a function of charge carrier concentration (not shown). This procedure to determine the relaxation time has been previously used in the literature as well [92,93].For comparison, we also show the evolution of the density of states with the chemical potential at 0 K. As discussed above and by others [40,42,43], the RBA works very well for these compounds and in the following we will discuss how the thermoelectric properties change with the charge carrier concentration within this approximation.

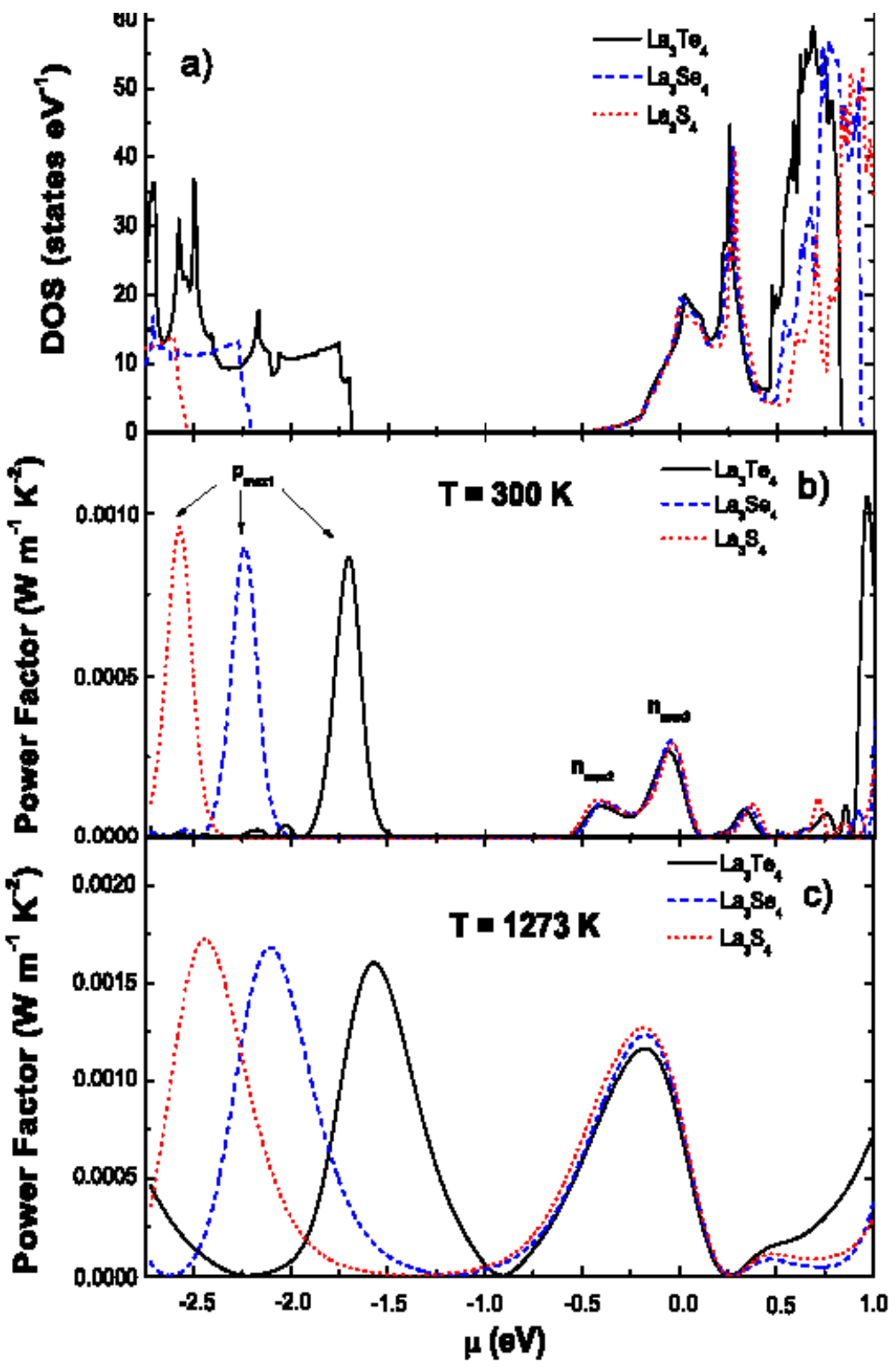

**Figure 7**

At 300 K, one can see maxima in the power factor for three different chemical potentials with one of them being located inside the valence band, whereas the two others are located in the conduction band. At 1273 K, there are only two maxima for the power factor– one in the conduction and the other in the valence band. The chemical potentials corresponding to these maxima are strongly shifted compared to the maxima at 300 K. This means that the best charge carrier concentration (corresponding to a maximum PF) is changing when the temperature is increasing. We will come back to this point later in more details.

As can be seen from Fig. 7, the PF is not very high for the stoichiometric compounds $La_3X_4$ and it is necessary to down-shift the chemical potential μ in order to improve their PF and hence their thermoelectric properties. For this purpose, it is essential to decrease the number of electrons in these compounds and - as the experiments have shown in the past - the introduction of vacancies is a very efficient way to do that.

In Figure 8a we report the temperature dependence of the Seebeck coefficient for $La_3Te_4$ and$La_3Se_4$ in the case of n-type conductivity for five different charge concentrations from the stoichiometric compound (n = 1 e/f. u.) down to a very low charge carrier concentration. Note that among the calculated

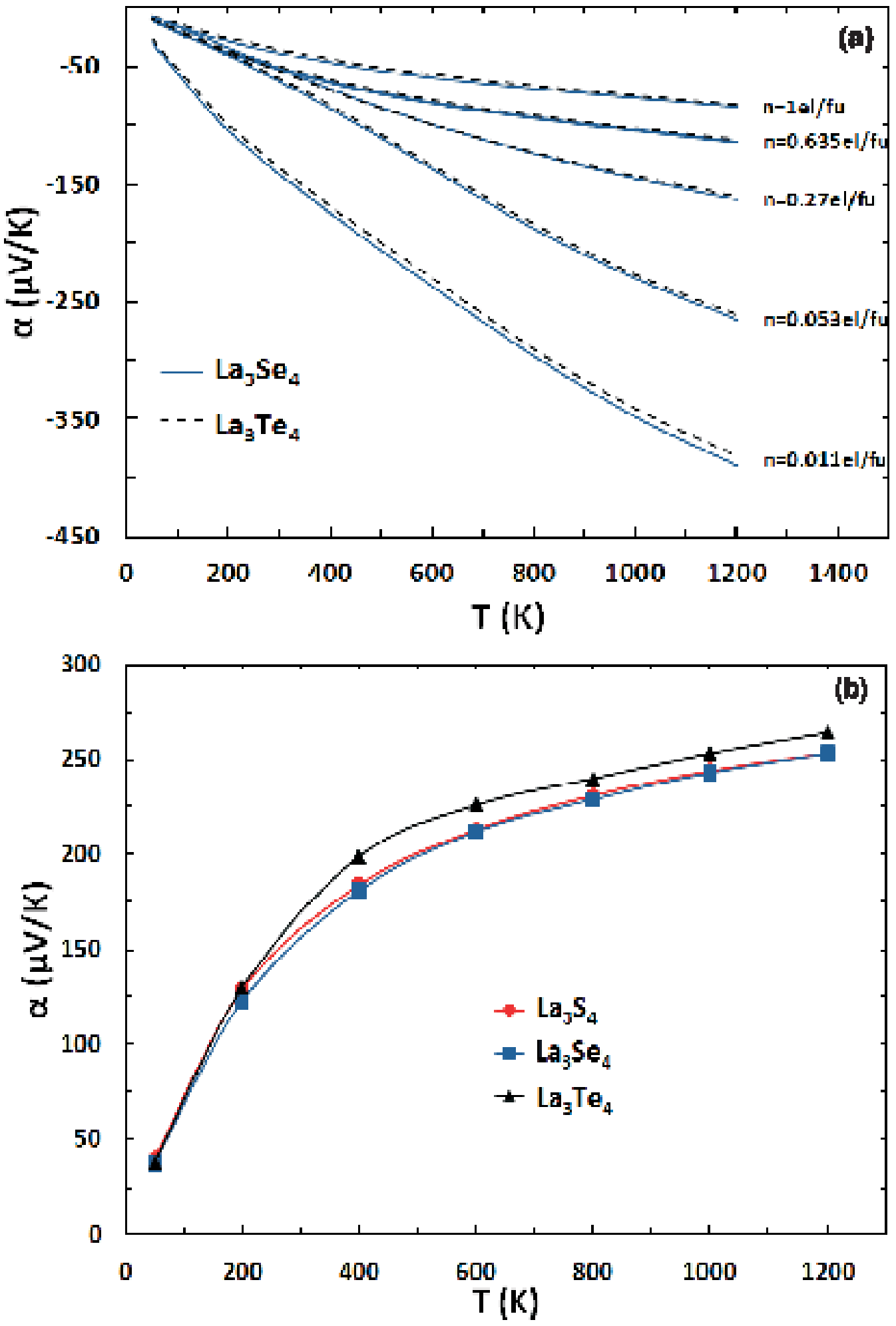

**Figure 8**

Seebeck coefficients reported in Fig. 8a, we have chosen to show the Seebeck coefficient for the charge carrier concentrations that correspond to or are close to the two maxima observed previously in the plot of the PF vs μ at 300 K ($n_{max3}$ = 0.63 e/f. u. and n $_{max2}$ = 0.011e/f. u.).In Figure 8b we show the temperature dependence of the Seebeck coefficient for all three compounds in the case of p-type conductivity for the charge carrier concentration corresponding to the PF maximum observed previously in the plot of the PF vs μ at 300K (Fig. 7):$p_{max1}$ = 0.117, 0.14 and 0.144 h/f. u. for respectively the telluride, selenide and the sulfide. The Seebeck coefficient for the telluride is the largest since it has the smallest charge carrier concentration whereas the selenide and the sulfide have similar Seebeck coefficients since the charge carrier concentrations are very close. One can see that the temperature dependence and the absolute values of the thermopower are very similar for all three compounds. This is due to their similar electronic band structure, especially in the case of the conduction band. For the n-type doping of $La_3Te_4$, our results are in very good agreement with May's results [40]. More importantly our results also qualitatively agree with the experiments.

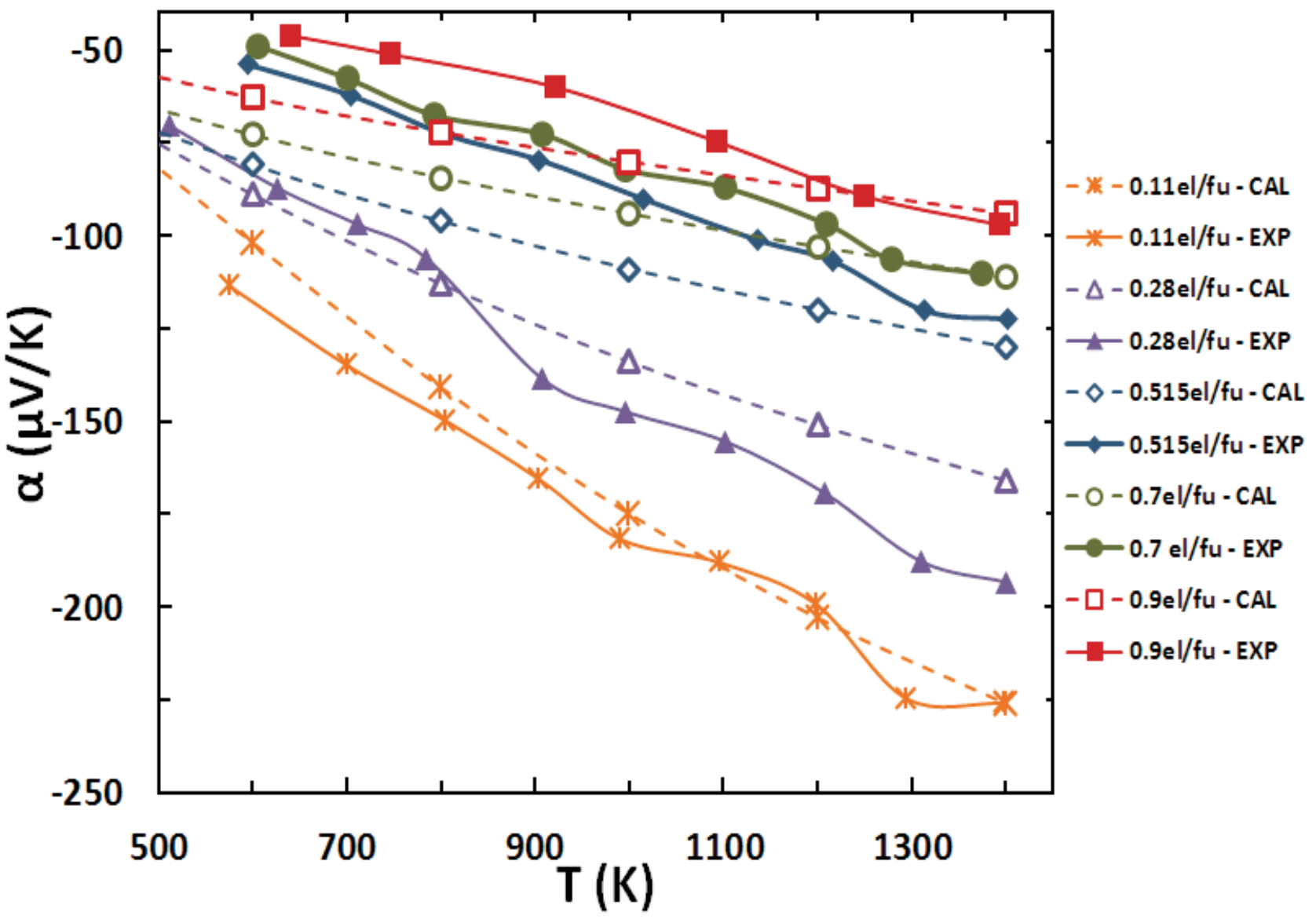

**Figure 9**

In Fig. 9, we compare our results for the Seebeck coefficient in the case of the sulfide with the experiments and more specifically with the data obtained by Wood et al. [30] since the whole set of experimental points has been obtained in the same conditions. Once more our results agree qualitatively with the experiments [30] but not fully quantitatively. We note globally a better agreement at high temperature than at room temperature. This observation also applies to the case of the telluride. We also note a better agreement at high temperature when the thermopower of the telluride and the sulfide is plotted as a function of the charge carrier concentration at 300 K and 1273 K (Fig. 10). It is clear that there is a good agreement between our calculations and the experiments at 1273 K [30,39] but the agreement is slightly less satisfactory at 300 K and 400 K [29,35,39]. Our results also agree very well with previous results of May et al at 400 K for the telluride [40]. However, for this last compound, both our calculations and May's calculations partly disagree with the available experiments at 400 K [39,40] because the Seebeck coefficient is higher than in the experiments above 0.5 electron/f. u. and lower than in the experiments below 0.5 electron/f. u..

In order to determine the PF, we need to calculate the electrical conductivity which is depending on the relaxation time τ. In order to evaluate realistically τ, as mentioned earlier, we have adjusted our calculated electrical conductivity on the experimental data at 1273 K. At this temperature, we have a complete set of data as a function of the charge carrier concentrations for both the sulfide and the telluride and more importantly, we can use the same relaxation time τ ($2*10^{-15}$ s) in order to fit our calculations to the experiments. Using this time relaxation, we can therefore calculate the power factor at 1273 K and at room temperature and we find a better agreement with the experiment at 1273 K (see Fig. 11), which is not surprising since we have the best agreement for the Seebeck coefficient between our calculations and the experiments at this temperature.

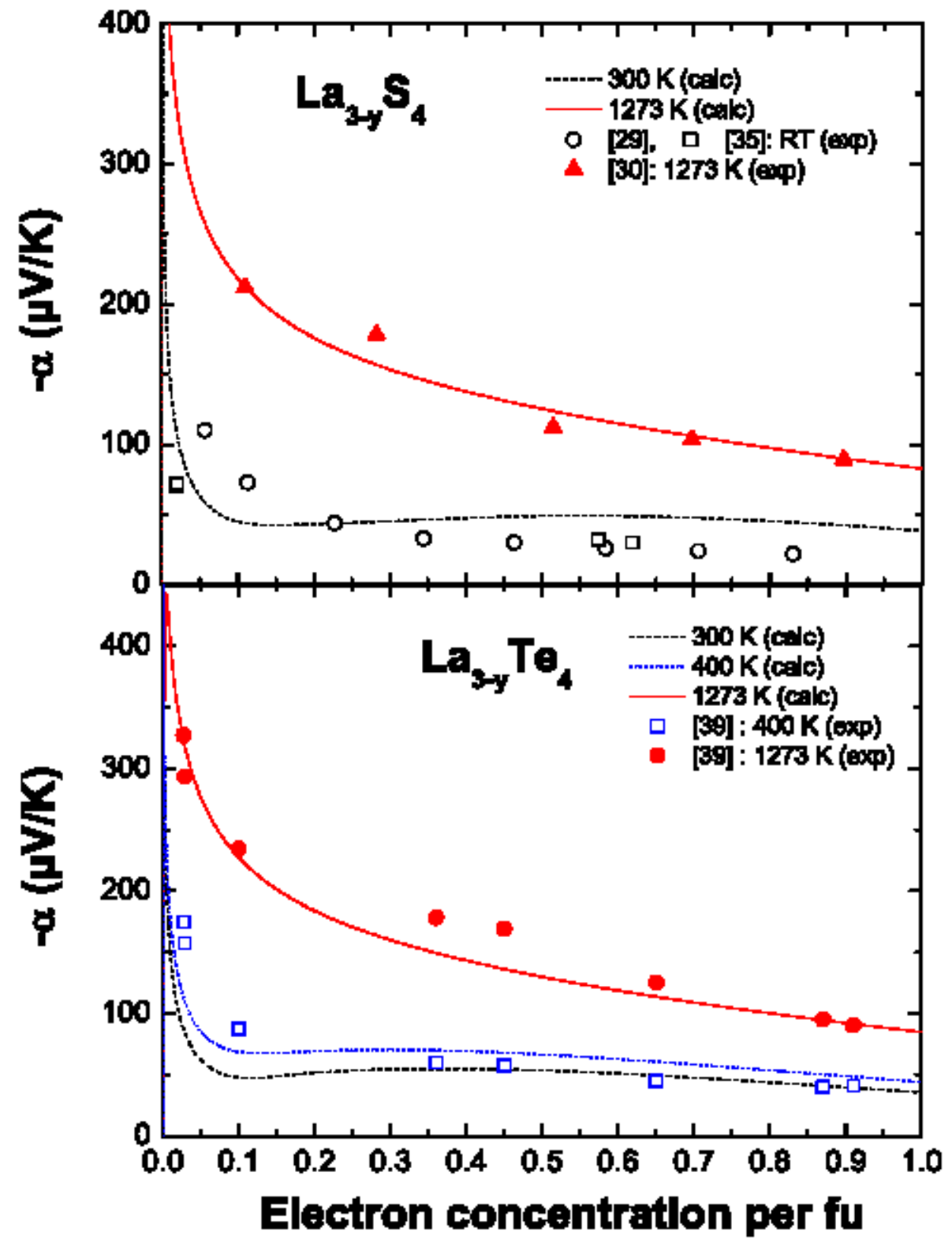

**Figure 10**

Although short, the relaxation time we have defined is inside the usual range of relaxation times typically found in the literature ($10^{-15}$ to $10^{-13}$ s) [92-94]. As expected, after analyzing how the PF varies with the chemical potential, we find that there are two electron concentrations and one hole concentration for which the PF goes through a maximum at 300 K. At 1273 K, there is one electron concentration and one hole concentration for which the PF is maximum. These maxima are at slightly different values of electron/hole concentration for the three different compounds. The values of these concentrations are reported in Fig. 11.Our results compare well with the experiments for $La_{3-y}Te_4$ since a maximum of the PF is located at about 0.5 electron per f.-u in this compound [39]. The comparison is not as good for the sulfide since the maximum found experimentally at 1273 K is at a much higher electron concentration (about 0.7 electron/f. u.) [30] and no clear tendency in the experimental data can be seen at room temperature [35,36].

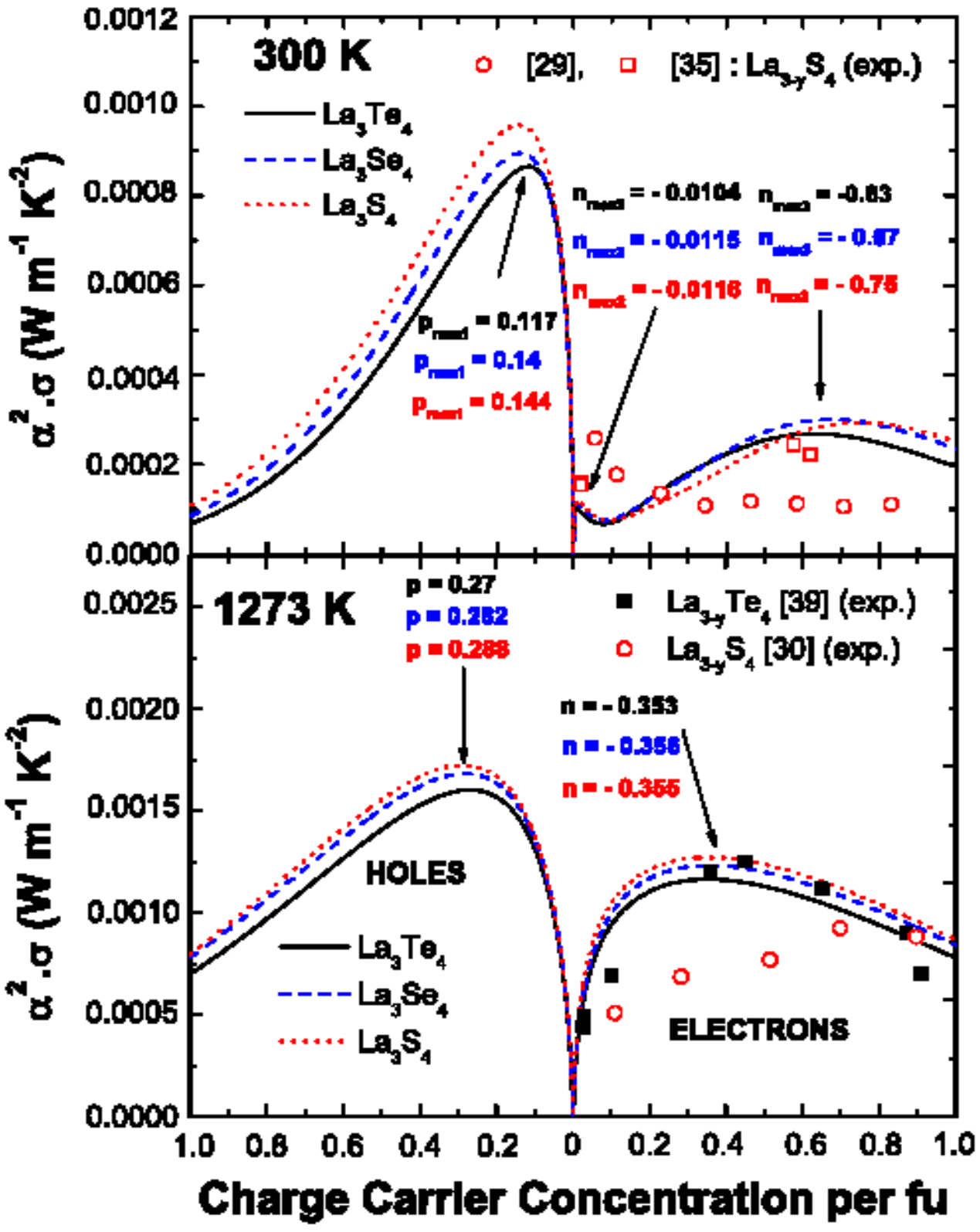

**Figure 11**

These calculations permit to predict that if we were able to p-dope the $La_3X_4$ compounds, we would get a slightly larger PF (at high temperature) and a much larger PF (at room temperature) than for the best n-doped $La_3X_4$ compounds(these predictions are independent of the choice of the relaxation time).This points out that this family of compounds has more promising thermoelectric properties than it was previously thought. However, as noted before, the charge carrier concentration to obtain the best TE properties is certainly lower than the charge carrier concentration for the best PF because of the large electronic thermal conductivity for compounds with such a large charge-carrier concentration. To verify this assertion, we have calculated the dimensionless Figure of Merit ZT with the assumption of $\kappa_l = 1$ W/m.K. This is a very reasonable assumption, as can be seen from the previous discussion on the lattice dynamics and thermal properties and such kind of calculations have already been done and give good results using the same approach [92-94]. The results are given in Fig. 12.

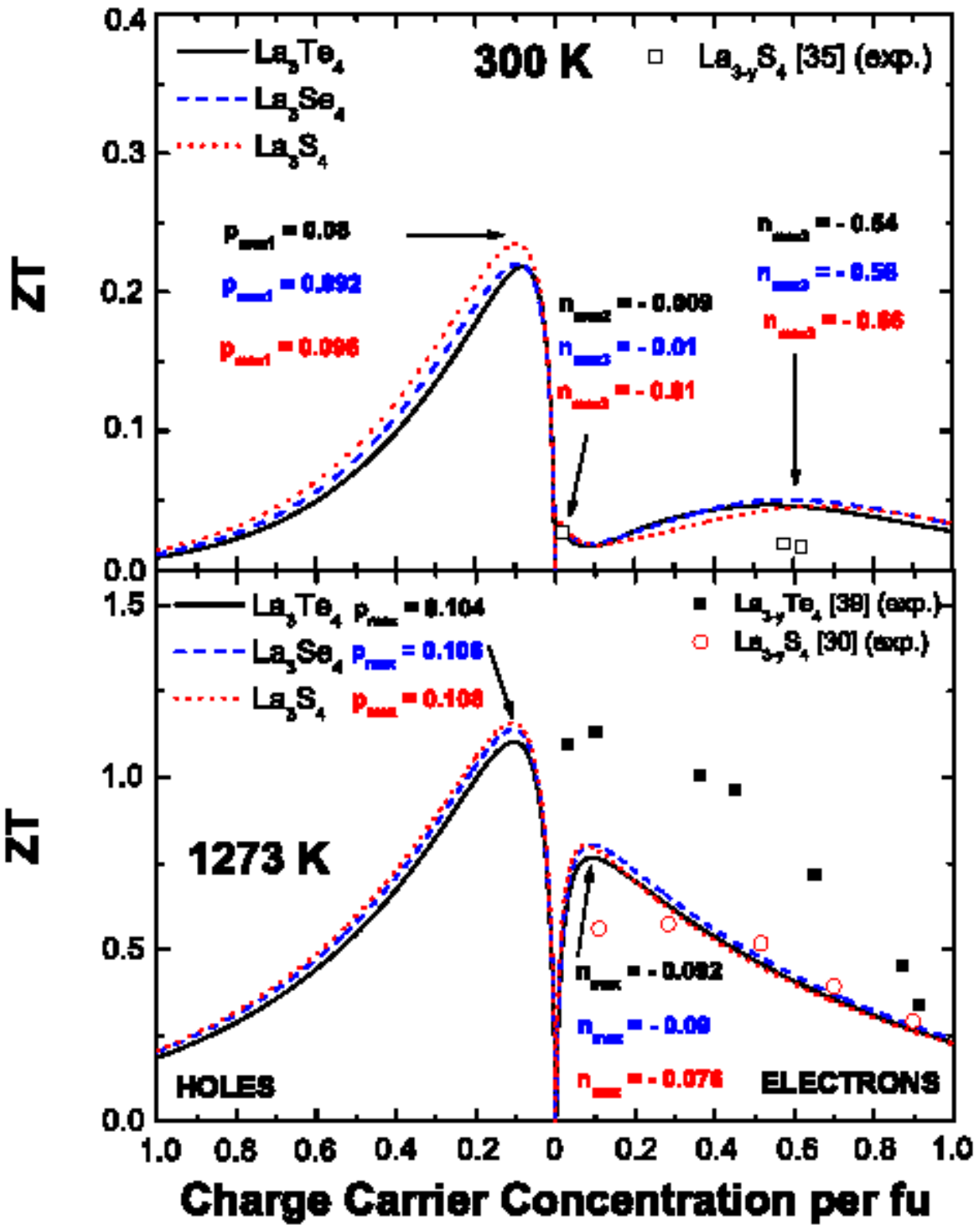

**Figure 12**

One can see that the maximum ZT is obtained for different (generally lower) charge carrier concentrations than for the PF. This is because of the strong contribution of the electron thermal conductivity that is larger than the lattice thermal conductivity for charge carrier concentrations higher than 0.16-0.2 charge carrier per f. u. for both n- and p-doping in these compounds. This has already been observed experimentally both in the sulfides [30] and tellurides [39]but for a slightly higher charge carrier concentration (about 0.3 electron/f. u.). At 1273 K, for the sulfide, we find a ZT slightly larger than in the experiments for n < 0.3 electron/f. u. and very close to the experiment above this value. In the case of the telluride, we find a ZT smaller than in the experiments in the full range of charge carrier concentrations. The main reason of this disagreement comes from our overestimated calculated thermal conductivity (including electronic and lattice parts) compared to May's experiments [39]. One can not exclude that the smaller thermal conductivity found in these experiments could be related to the nanometric size of the grains. Another explanation of this disagreement could come from a limitation of the use of the Wiedemann-Franz law, since we use this law for calculating the electronic thermal conductivity in all the calculations. From our calculations at 1273 K, for n-doping, one can see that the maximum ZT is around 0.08-0.09 electron/f. u., a value slightly lower than in the experiments (about 0.1 electron/f. u.). From our calculations, for p-type compounds, one predicts a relatively large ZT approaching 0.2-0.25 at room temperature for all three compounds for 0.08-0.1 hole/f. u. and a ZT as large as 1.1-1.2 at 1273 K for a slightly larger hole concentration of about 0.1-0.11 hole/f. u. It is important to note that we find a ZT 4-5 times larger for p-type than for n-type compounds at room temperature and a ZT about 50 % larger for p-type than for n-type compounds at 1273 K. As the largest ZT found for the $La_3X_4$ compounds at this temperature is about 1.15 for n-type compounds, this means that a ZT between 1.5 and 2 could be reached for p-type $La_3X_4$ compounds.

How can one get a p-doped $La_3X_4$ compound? It is possible to substitute one-third of the lanthanum atoms by some alkaline-earth element A, to get semiconductors of $ALa_2X_4$ composition, especially for the sulfides [6,15, 36,74]. Starting from such a compound, it could be possible to obtain p-doped semiconductors if one substitutes the chalcogen atoms by an element of the fifth column. If such doping was possible and if the band structure of the $ALa_2X_4$ semiconductors is still similar to the one of the $La_3X_4$ compounds, then it could be possible to obtain new highly-efficient thermoelectric compounds of p-type. Another possibility would be to substitute the rare-earth element by an alkaline metal M such as sodium. It has been shown recently that this is possible in $M_zLa_{2-z}X_3$[95]. However, it could be difficult to stabilize high doping levels this way.

## 4. Conclusion

In the present paper, we have reported ab-initio calculations of the structural stability, lattice dynamics, electronic and thermoelectric properties of the lanthanum chalcogenides$\gamma$-$La_3X_4$ ($X$=S,Se,Te) with cubic symmetry in order to evaluate their potential for future applications in high-temperature electrical thermogeneration. We have found large formation energies explaining their refractory properties. We have shown that the lanthanum motions are strongly coupled with the tellurium motions in $La_3Te_4$, whereas the lanthanum motions are strongly decoupled with the sulfur motions in $La_3S_4$. Despite the strong difference of the coupling between the chalcogen and the lanthanum atoms and the absence of cages in the $Th_3P_4$ structure, we have shown that the Umklapp scattering of the acoustical phonons, notably by the low-energy optical modes, are able to explain their intrinsically low lattice thermal conductivity. If the vacancies can reduce further the lattice thermal conductivity, their main impact on the thermal conductivity is to reduce strongly the electronic thermal conductivity. Indeed, as confirmed by our calculations, the presence of vacancies reduces the electron concentration and when this concentration y approaches 0.33 in $La_{3-y}Te_4$, the Fermi level shifts down inside a wide energy bandgap making the compound semiconducting. We show that the electronic structure of all three compounds is very similar and that the width of the energy bandgap increases from the telluride to the sulfide, in good agreement with experiments. For this reason, all three compounds have similar thermoelectric properties. Our results therefore show that similar thermoelectric properties could be reached in the sulfides and selenides than in the tellurides, with the advantage of a much higher abundance of the sulfur and selenium compared to the tellurium and their lower toxicity. We show that the best charge carrier concentration for a maximum power factor for both n- and p-type doping is different from the one for optimizing the ZT, mainly because of the decrease of the electronic thermal conductivity with charge carrier concentration. We predict better thermoelectric properties for p-type $La_{3-y}X_4$ than for n-type $La_{3-y}X_4$, depending of the temperature range. We also predict a ZT reaching values as large as at least 0.25 and 1.5 at respectively 300 K and 1273 K if these compounds could be optimally p-doped. Clearly, the sulfide compound, $La_{3-y}S_4$, has a very high potential for high temperature thermoelectric applications because of its potentially large ZT for both n- and p- type if adequately doped and because it is made of abundant and relatively cheap elements.

**Acknowledgments:** We thank the computer centers CINES and HPC@LR in Montpellier for their support. We acknowledge the financial support of the CNRS through the “Programme Interdisciplinaire Energie”. We thank Total and Hutchinson for supporting financially the research work of K.N. and P.J.
R.V. and K. N. have equally contributed to this work.

## Table Captions

Table 1 Calculated structural parameters, bulk modulus B and its pressure derivative B' and formation enthalpies of $La_3X_4$ compounds, compared with experiments [22,23,31,42,62,63,66].

Table 2 Experimental structure parameters and formation enthalpies of $La_2X_3$ compounds as found in the literature [10,12,13,64,65,67]

Table 3 Energies of the calculated vibrational modes at Γ point (in meV)

Table 4 Energies of the Raman-active vibrational modes as determined experimentally (in meV) [73-76]. When the $T_2$ label is followed by a ?, it means that this is a tentative assignments made by the authors of ref. 77.

Table 5 Calculated averaged anisotropic atomic displacement parameters $U_{ij} = \langle u_i u_j \rangle$ (i, j = x, y, z) and isotropic $U_{iso}$ of lanthanum and chalcogen *X* atoms for the $La_3X_4$ compounds

Table 6 Experimental values of averaged anisotropic atomic displacement parameters $U_{ij} = \langle u_i u_j \rangle$ (i, j = x, y, z) and isotropic $U_{iso}$ of lanthanum and chalcogen *X* atoms for the $La_2X_3$ compounds as found in the literature [10,12,13]

Table 7 Debye temperatures and room temperature specific heat of$La_3X_4$ compounds compared to the experimental values [8,14,41,64]

Table 8 Sommerfeld coefficients of $La_3X_4$ compounds compared to the experimental values [14,41]

Table 9 Bader charges of the atoms in the different $La_3X_4$ compounds

## Figure Captions

Fig. 1 Phonon dispersion curves (different colors correspond to different symmetries: green = $A_1$, olive = $A_2$, red = $E$, blue = $T_1$, black = $T_2$), total and partial phonon density of states and cumulated spectral weight (CSW) of $La_3Te_4$.

Fig. 2 Phonon dispersion curves (different colors correspond to different symmetries: see Fig. 1), total and partial phonon density of states and cumulated spectral weight (CSW) of $La_3Se_4$

Fig. 3 Phonon dispersion curves (different colors correspond to different symmetries: see Fig. 1), total and partial phonon density of states and cumulated spectral weight (CSW) of $La_3S_4$

Fig. 4 Calculated heat capacity compared to previously published experimental results [41] for $La_3Te_4$. Inset: plot of the calculated $C_V/T^3$vs T

Fig. 5 Electronic band structure of $La_3X_4$compounds

Fig. 6 (a) Electronic density of states of $La_3Te_4$; (b) Electronic density of states of $La_{11}Te_{16}$

Fig. 7 (a) Electronic density of states vs chemical potential in $La_3X_4$ ;(b) Power factor vs chemical potential in $La_3X_4$at 300 K ; (c) Power factor vs chemical potential in $La_3X_4$ at 1273 K

Fig. 8(a) Calculated thermoelectric power for $La_{3-y}X_4$ ($X$ = Se, Te) vs temperature for different charge carrier concentrations; (b) Calculated thermoelectric power for $La_{3-y}X_4$ ($X$ = S, Se, Te) vs temperature for the charge carrier concentration corresponding to the first maximum $p_{max1}$ of the PF for p-doping.

Fig. 9 Calculated Seebeck coefficient for $La_{3-y}S_4$vs temperature for different charge carrier concentrations compared to experiments. Charge carrier concentrations are identical in calculations (open symbols, dashed lines) and experiments (full symbols, full lines) [30].

Fig. 10 Calculated thermoelectric power for $La_{3-y}X_4$ ($X$ = S, Te) vs electron concentration at 300 K, 400 K (for the sulfide) and 1273 K compared to experiments [29,30,35,39].

Fig. 11 Calculated PF for $La_{3-y}X_4$ ($X$ = S, Se, Te) vs charge carrier concentration at 300 K and 1273 K compared to experiments [29,30,35,39].

Fig. 12 Calculated ZT for $La_{3-y}X_4$ ($X$ = S, Se, Te) vs charge carrier concentration at 300 K and 1273 K compared to experiments [29,30,35,39].

**Tables**

Table 1

| Compound | a (Å) | $x_X$ | B (Gpa) | B' | Formation enthalpy (ΔH) eV/atom |
|---|---|---|---|---|---|
| $La_3S_4$ | 8.74 | 0.075115 | 73.8 | 5.2 | -2.276 |
| | 8.727-8.73 [23,31] | - | 71.3 [66] | 5.3 [66] | - |
| $La_3Se_4$ | 9.0982 | 0.07504 | 63 | 4.8 | -2.038 |
| | 9.049-9.055 [63] | 0.075 [62] | - | - | -(1.984-2.025) [63] |
| $La_3Te_4$ | 9.6874 | 0.07513 | 50.5 | 4.8 | -1.621 |
| | 9.628-9.634 [22,42] | - | 50 [39] | - | - |

Table 2

| Compound | a (Å) | $x_X$ | Formation enthalpy (ΔH) eV/atom |
|---|---|---|---|
| $La_2S_3$ | 8.731 [11] | 0.0734 [11] | -2.446 [65], -2.5066 [67] |
| $La_2Se_3$ | 9.0521 [13] | 0.07392 [13] | -1.934 [64,67] |
| $La_2Te_3$ | 9.619 [14] | 0.0748 [14] | -(1.5-1.626) [67] |

Table 3

| Compounds | $A_1$ modes | $A_2$ modes | E modes | $T_1$ modes | $T_2$ modes |
|---|---|---|---|---|---|
| $La_3S_4$ | 23.71 | 13.17, 28.79 | 7.27, 29.43, 32.21 | 11.01, 13.92, 20.95, 24.48, 31.03 | 7.91, 11.31, 19.205, 22.87, 27.38 |
| $La_3Se_4$ | 14.48 | 11.48, 18.1 | 6.77, 18.23, 19.765 | 9.25, 11.16, 14.455, 16.89, 19.19 | 7.74, 9.62, 13.49, 16.89, 17.21 |
| $La_3Te_4$ | 10.95 | 9.08, 14.29 | 6.51, 13.93, 15.06 | 7.7, 8.405, 12.12, 14.33, 14.99 | 7.78, 8.38, 11.65, 13.5, 14.73 |

Table 4

| Compounds | Raman modes (meV) |
|---|---|
| $La_3S_4$ | No data |
| $La_2S_3$ | 9.7 (E), 10.8 ($T_2$), 19.8 ($T_2$ ?), 21.7 ($T_2$), 22.9 ($A_1$), 28.1 (E), 33.5 ($T_2$ ?), 35.1 ($T_2$), 37.2 (E) [76]<br>10.5, 15.5, 23.1, 28.4, 34.7 [77]<br>13±1.5, 23±1, 26.5±0.5 ($T_2$ TO modes from IR experiments) [78] |
| $La_3Se_4$ | 6.9, 16.1, 19.8, 21.6, 30 [75] |
| $La_3Te_4$ | No data |

Table 5

| Atom type | $U_{iso}$ (Å$^2$) | $U_{xx}$ (Å$^2$) | $U_{yy}$ (Å$^2$) | $U_{zz}$ (Å$^2$) | $U_{yz}$ ($10^{-4}$Å$^2$) | $U_{zx}$ ($10^{-4}$Å$^2$) | $U_{xy}$ ($10^{-4}$Å$^2$) |
|---|---|---|---|---|---|---|---|
| $La_3Te_4$ | | | | | | | |
| La | 0.01268 | 0.01253 | 0.01274 | 0.012765 | 0.167 | 0.0883 | -0.867 |
| Te | 0.0109 | 0.01075 | 0.01095 | 0.01098 | 0.179 | 0.13 | -0.814 |
| $La_3Se_4$ | | | | | | | |
| La | 0.01115 | 0.0111 | 0.01118 | 0.01117 | 0.115 | -0.15 | -0.0833 |
| Se | 0.01025 | 0.01019 | 0.01028 | 0.01027 | 0.1025 | -0.146 | -0.091 |
| $La_3S_4$ | | | | | | | |
| La | 0.010165 | 0.0101 | 0.01019 | 0.0102 | 0.2 | -0.13 | -0.313 |
| S | 0.01 | 0.0099 | 0.01 | 0.01 | 0.169 | -0.076 | -0.338 |

Table 6

| Atom type | $U_{iso}$ (Å$^2$) | $U_{xx}$ (Å$^2$) | $U_{yy}$ (Å$^2$) | $U_{zz}$ (Å$^2$) | $U_{yz}$ ($10^{-4}$Å$^2$) | $U_{zx}$ ($10^{-4}$Å$^2$) | $U_{xy}$ ($10^{-4}$Å$^2$) |
|---|---|---|---|---|---|---|---|
| $La_2Te_3$ [14] | | | | | | | |
| La | 0.0152 | | | | | | |
| Te | 0.01216 | | | | | | |
| $La_2Se_3$ [13] | | | | | | | |
| La | 0.0117 | 0.0138 | 0.0107 | 0.0107 | 0 | 0 | 0 |
| Se | 0.0107 | 0.0107 | 0.0107 | 0.0107 | 1 | 1 | 1 |
| $La_2S_3$ [11] | | | | | | | |
| La | 0.0156 | 0.0166 | 0.0157 | 0.0157 | 0 | 0 | 0 |
| S | 0.0138 | 0.0138 | 0.0138 | 0.01 | 0.5 | 0.5 | 0.5 |

Table 7

| Compound | $\Theta_D{}^C$ (K) from heat capacity | $\Theta_D{}^{int}$ (K) from integration | heat capacity at 298 K (in $k_B$/at) | Remarks |
|---|---|---|---|---|
| $La_3S_4$ | 263.7 | 290.9 | 2.847 | This work |
| | 227 [14] | | 3.019 [8] | Exper. |
| $La_3Se_4$ | 228.1 | 204.5 | 2.9277 | This work |
| | 201 [14] | | 3.134 [64] | Exper. |
| $La_3Te_4$ | 198.9 | 168.5 | 2.9513 | This work |
| | 184 [41], 205 [14] | 173 [41] | 3.006 [41] | Exper. |

Table 8

| Compound | $\gamma$ (mJ/mole.K$^2$) | Remarks |
|---|---|---|
| $La_3S_4$ | 2.98 | This work |
| | 3.09 [14] | Exper. |
| $La_{2.973}S_4$* | 3.015 | This work |
| | 3.67 [14] | Exper. |
| $La_3Se_4$ | 3.21 | This work |
| | 2.57 [14] | Exper. |
| $La_{2.985}Se_4$* | 3.02 | This work |
| | 3.59 [14] | Exper. |
| $La_3Te_4$ | 2.56 | This work |
| | 1.93 [14], 2.91 [41] | Exper. |

*Using the shifted Fermi level obtained from the RBA with respect to the Fermi level of $La_3X_4$.

Table 9

| Atoms | $La_3S_4$ | $La_3Se_4$ | $La_3Te_4$ |
|---|---|---|---|
| La | +1.7031 | +1.6287 | +1.5216 |
| S | -1.2773 | | |
| Se | | -1.22155 | |
| Te | | | -1.1412 |